\documentclass[fleqn,usenatbib]{mnras}

\usepackage{newtxtext,newtxmath}

\usepackage[T1]{fontenc}

\DeclareRobustCommand{\VAN}[3]{#2}
\let\VANthebibliography\thebibliography
\def\thebibliography{\DeclareRobustCommand{\VAN}[3]{##3}\VANthebibliography}

\usepackage{graphicx}	
\usepackage{amsmath}	
\usepackage{xcolor}
\usepackage{txfonts}
\usepackage{url}
\UseRawInputEncoding

\title[Expert visual inspection of strong lenses]{The impact of human expert visual inspection on the discovery of strong gravitational lenses}

\author[K. Rojas et al.]{\parbox{\textwidth}{
Karina Rojas,$^{1}$\thanks{karina.rojas@port.ac.uk}
Thomas E. Collett,$^{1}$ \thanks{thomas.collett@port.ac.uk}
Daniel Ballard,$^{1}$
Mark R. Magee,$^{1}$
Simon Birrer,$^{2,3,4}$
Elizabeth Buckley-Geer.$^{5,6}$
James~H.~H.~Chan,$^{7}$ $^{8}$
Benjamin Cl\'ement,$^{9}$
Jos\'e M. Diego,$^{10}$
Fabrizio Gentile,$^{11,12}$
Jimena Gonz\'alez,$^{13}$
R\'emy Joseph,$^{14}$
Jorge Mastache,$^{15,16}$
Stefan Schuldt,$^{17,18}$
Crescenzo Tortora,$^{19}$
Tom\'as Verdugo,$^{20}$
Aprajita Verma,$^{21}$
Tansu Daylan,$^{22,23}$
Martin Millon,$^{9}$
Neal Jackson,$^{24}$
Simon Dye,$^{25}$
Alejandra Melo,$^{17}$
Guillaume Mahler,$^{26,27}$
Ricardo L. C. Ogando,$^{28}$
Fr\'ed\'eric Courbin,$^{9}$
Alexander Fritz,$^{29}$
Aniruddh Herle,$^{17,30}$
Javier A. Acevedo Barroso,$^{9}$
Raoul Ca\~nameras,$^{17}$
Claude Cornen,$^{31}$
Birendra Dhanasingham,$^{32}$
Karl Glazebrook,$^{33}$
Michael N. Martinez,$^{13}$
Dan Ryczanowski,$^{34}$
Elodie Savary,$^{9}$
Filipe G\'ois-Silva,$^{28}$
L. Arturo Ure\~na-L\'opez,$^{35}$
Matthew P. Wiesner,$^{36}$
Joshua Wilde,$^{37}$
Gabriel Valim Cal\c{c}ada,$^{28}$
R\'emi Cabanac,$^{38}$
Yue Pan,$^{6}$
Isaac Sierra,$^{6}$
Giulia Despali,$^{39}$
Micaele~V.~Cavalcante-Gomes,$^{40,28}$
Christine Macmillan,$^{30}$
Jacob Maresca,$^{25}$
Aleksandra Grudskaia,$^{17}$
Jackson H. O'Donnell,$^{41,42}$
Eric Paic,$^{9}$
Anna Niemiec,$^{26,27}$
Lucia F. de la Bella,$^{1}$
Jane Bromley,$^{42}$
Devon M. Williams,$^{43}$
Anupreeta More,$^{44,45}$
Benjamin C. Levine.$^{4}$
\newline
\emph{\normalsize Affiliations are listed at the end of the paper}
}}

\date{Accepted XXX. Received YYY; in original form ZZZ}

\pubyear{2022}
\begin{document}
\label{firstpage}
\pagerange{\pageref{firstpage}--\pageref{lastpage}}
\maketitle
\begin{abstract}
We investigate the ability of human 'expert' classifiers to identify strong gravitational lens candidates in Dark Energy Survey like imaging. We recruited a total of  55 people that completed more than 25$\%$ of the project. During the classification task, we present to the participants 1489 images. The sample contains a variety of data including lens simulations, real lenses, non-lens examples, and unlabeled data. We find that experts are extremely good at finding bright, well-resolved Einstein rings, whilst arcs with $g$-band signal-to-noise less than $\sim$25 or Einstein radii less than $\sim$1.2 times the seeing are rarely recovered. Very few non-lenses are scored highly. There is substantial variation in the performance of individual classifiers, but they do not appear to depend on the classifier's experience, confidence or academic position. These variations can be mitigated with a team of 6 or more independent classifiers. Our results give confidence that humans are a reliable pruning step for lens candidates, providing pure and quantifiably complete samples for follow-up studies.

\end{abstract}


\newcommand{\tc}[1]{{\color{green}{\sf{[TC: #1]}}}}
\newcommand{\red}[1]{{\color{red}{#1}}}



\section{Introduction}

The phenomenon of strong gravitational lensing has enormous power as a tool to study a variety of cosmological questions. For example, strong lenses enable a magnified view of the high redshift Universe \citep{Christensen2012,Stark2015,ShuY2016,Ebeling2018,ShuX2022}, a direct probe of dark matter in galaxies, clusters and substructures \citep{Oguri2002,Vegetti2010,Jimenez-Vicente2015a,Nierenberg2017,Gilman2020} and a geometrical probe of the cosmological parameters \citep{Collett-Auger2014,Bonvin2017,Wong2020}. Despite the potential of this tool, almost all applications of strong lensing are limited by sample size, but the era of deep wide area surveys offers an opportunity to grow strong lens samples a hundredfold \citep{Collett2015}, and with that improve current studies and enable the exploration of novel ideas as lens supernovae or compounds lenses.

As astronomy enters the era of billion object surveys, sophisticated methods for discovering strong lenses have been developed \citep[e.g.][]{Lanusse2018,Avestruz2019} and applied \citep{Jacobs2017,Jacobs2019A,Jacobs2019B,Rojas+2022,Savary+22,Petrillo2017,Petrillo2019}. These methods have been extremely successful at identifying candidate lenses, but the rarity of lenses means that even classifiers with 99.99\% accuracy produce 100 false positives for every true lens. The gold standard for confirming a lens is spectroscopic confirmation of multiple redshifts, but reducing false positive rates is needed for a spectroscopic confirmation campaign to be viable. In most cases, a human expert step is used as the final stage filtering step to remove false positives.

Expert human classification has been very successful at identifying the best strong lens candidates. For example, \citet{Tran+2022} recently targeted 79 lens candidates, spectroscopically confirming 53 and definitely ruling out only 4 \footnote{20 of the remaining objects are likely strong lenses but redshifts weren't obtained for both lens and source in 20 of the systems. 2 systems yielded no redshifts.}.  However, introducing a human into any classification task is likely to bring in selection biases: it is much easier to identify a bright arc that is well resolved from the lensing galaxy and significantly different in colour. 

The primary purpose of this work is to calibrate and understand how introducing human experts biases lens searches. In addition, we aim to understand how the choice of "experts" impacts the lens candidate sample selected and how search teams can mitigate the biases of their members. We set out to answer the following questions:
\begin{itemize}
\item What are the properties of lensing systems that human experts identify reliably as lenses, and what do they miss?
\item Do human experts confuse non-lenses for lenses?
\item How does expert classification depend on the experience and confidence of the experts?
\item How reliable are individual classifications?
\item When ranking lens candidates, what do the scores of teams of experts mean?
\item How should lens searchers best build an expert team to classify their candidates?
\end{itemize}

In Section~\ref{sec:experiment} we will lay out our experiment, data and expert participants. In Section~\ref{sec:resultsA} we investigate how subsets of the lens candidates are scored by the ensemble of our users, enabling us to understand the selection biases of our experts. In Section~\ref{sec:resultsB} we investigate how individual users perform on the classification task. We summarize the conclusions of this work in Section~\ref{sec:conclusions}.

\section{The experiment}
\label{sec:experiment}

Our experiment is designed to understand human expert biases when performing visual inspection of strong lens candidates. With experts we refer to any person involved in strong lensing research, with an academic status from masters student to Professor (or similar). Additionally, we invited a small number of citizen scientists from outside the academic strong lensing community. Several of these are experienced users from the Spacewarps project.  The details of our invitation to be part of this experiment can be found in Appendix~\ref{appendix-invitation}.

We used the citizen science web portal {\tt Zoouniverse}\footnote{\url{https://www.zooniverse.org/projects/krojas26/experts-visual-inspection-experiment}} to serve a sample of 1489 images 
for classification. 

The users were asked to choose the best description for the object displayed from four options: (1) Certain lens ($>90\%$), (2) Probable lens ($50-90\%$), (3) Probably not a lens ($2-50\%$), and (4) Very unlikely ($<1\%$). We included percentages of confidence to be a lens to avoid semantic uncertainty \footnote{Formally, these percentages cannot be probabilities since we did not provide the users with prior information on the composition of the sample being classified.}. 

We designed our experiment to closely mimic a real strong lens classification task. In real lens searches, there are no tutorials, and there is limited prior knowledge of the completeness and purity of the sample to be classified. We therefore avoided offering further guidelines about the composition of the data sets to be classified. For the same reason, we did not offer a tutorial nor examples as would be usual in a citizen science project.

\subsection{The data}
\label{sec:data-sets}

The image cutouts are from the Dark Energy Survey (DES). DES uses the Blanco 4-m telescope and the Dark Energy Camera \citep[DECam,][]{Honscheid2008,Flaugher2015} located at Cerro Tololo Inter-American Observatory (CTIO), Chile. The observations are performed in the optical $grizY$ bands. We used $gri$-bands to produce colour composite $50 \times 50$ pixels ($\sim 13 \arcsec \times 13 \arcsec$) images centering the object of interest in the middle. The images have a typical 5$\sigma$ depth of 23.72, 23.35, 22.88 in  $gri$ respectively.

We simultaneously display three different colour scalings of each image to facilitate the recognition of the different features in the stamp (see Fig.~\ref{fig:3scales}). The three imaging scalings are: Default, blue and sqrt, and they are described in Appendix~\ref{appendix-images}.

\begin{figure}
	\includegraphics[width=\columnwidth]{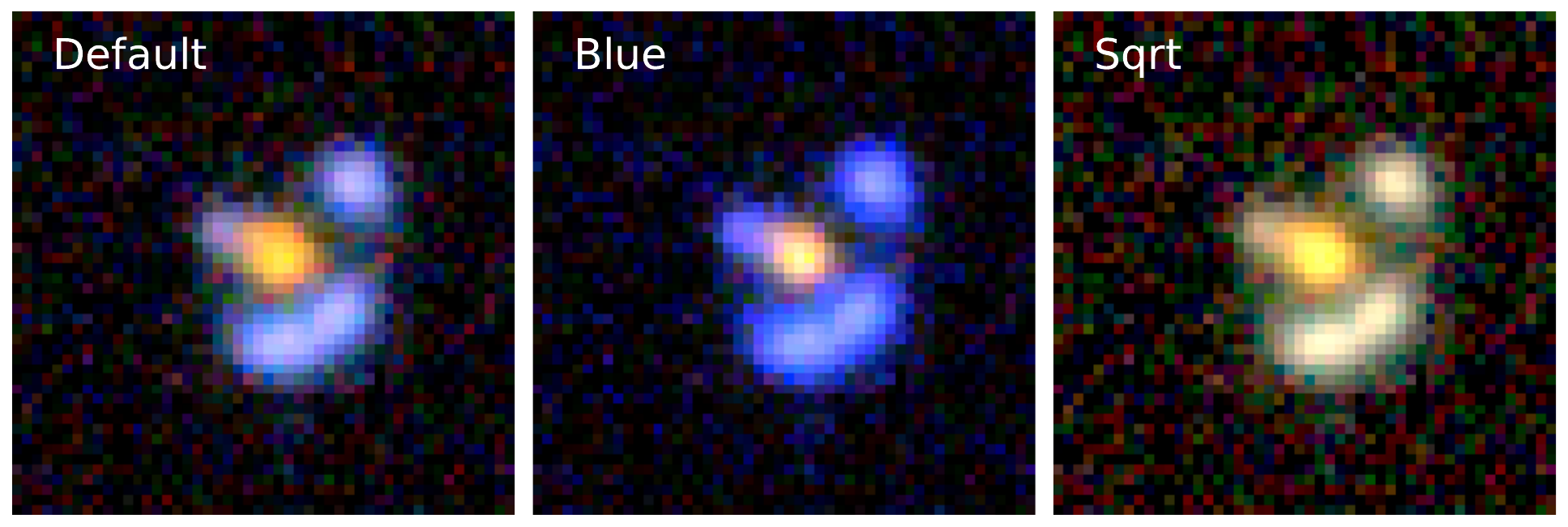}
    \caption{Example of a DES cutout of $50\times50$ pixels $(13\arcsec \times 13\arcsec)$ of an object displayed in the three different scales presented to the users in the experiment.}
    \label{fig:3scales}
\end{figure}

Since we aimed to assess the classification skills of experts, we required both a sufficiently large number of experts to participate and sufficient classifications per expert. This creates tension for experiment design since experts have difficulty engaging with the experiment if the classification task is too large. We decided that a sample size of around 1000 objects (around an hour of time assuming 2 seconds per classification) was small enough to get significant engagement and large enough to provide useful data.

Given the rarity of real lenses on the sky, a random sample of $\sim$1000 objects will not provide useful data. Instead, we designed a sample to have a broad variety of data including: simulated lenses; real lens candidates; non-lens examples; and unlabeled data. We also duplicated 105 cutouts, 15 cutouts from seven out of nine data sets excluding the eleven examples from SLACS and the four lenses from \cite{Rojas+2022}. The idea is to investigate the level of consistency of individuals when classifying. In total we had 1489 images to classify. 

\begin{figure}
	\includegraphics[width=\columnwidth]{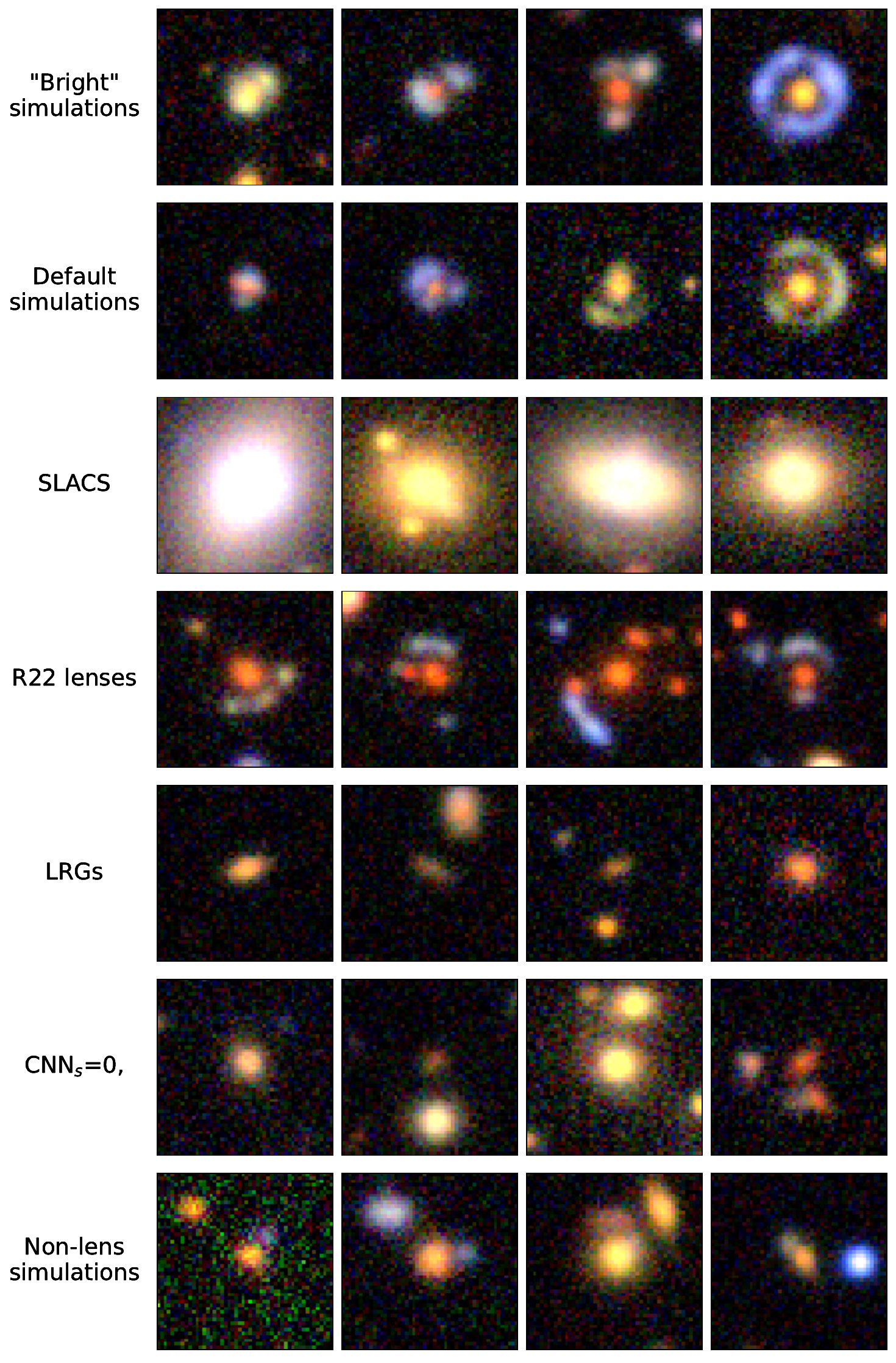}
            \caption{Example of objects presented in the different labelled data sets."Bright" simulations, Default simulations, SLACS and R22 lenses from \protect\cite{Rojas+2022} are examples of objects labelled as lenses, while LRGs, CNN$_{s}$=0 and Non-lens simulations are examples of objects labelled as non-lenses. The cutouts are $50\times50$ pixels and they are displayed in the default scale. }
    \label{fig:3scalesexamples}
\end{figure}

Below we describe the different data sets, which are in 3 main categories: lens simulations and lens candidates that are labelled as lenses; negative examples that are labelled as non-lenses; and unlabeled data. In Tab.~\ref{tab:datasets} we present the name, number of objects and category of each data set and in Fig.~\ref{fig:3scalesexamples} we present as an example the image of four objects of each labelled data set, more details about these sets can be found in the following sections. Our sample is not actively selected to include common false positives such as spiral, irregular, interacting or ring galaxies, it also does not include lensing by disk galaxies or groups of galaxies. This is due to the sample size of ~1500 objects.

\begin{table}
	\centering
	\caption{Data sets presented in the experiment.}
	\label{tab:datasets}
	\begin{tabular}{|c|c|c} 
		\hline
		Data set name &  number of objects & category \\
        \hline
		"Bright" simulations & 150 & Lens data set \\
		Default simulations  & 150 & Lens data set \\
		SLACS                & 11  & Lens data set \\
		R22 lenses           & 4   & Lens data set \\
		LRGs                 & 150 & Non-lens data set \\
		CNN$_{s}$ = 0              & 150 & Non-lens data set \\
		Non-lens simulations & 150 & Non-lens data set \\
		CNN-best             & 300 & Unlabeled data \\
		Random               & 300 & Unlabeled data \\
        \hline
	\end{tabular}
\end{table}

\subsubsection{Lens simulations and additional lens examples.}
We created two sets of simulations of strong lens systems. Each of them contains 160 images and both have a uniform Einstein radius distribution between $0.8 \arcsec < \theta_E < 3.0 \arcsec$.

The simulations were created using {\tt Lenstronomy}\footnote{\url{https://github.com/lenstronomy/lenstronomy}} \citep{Birrer2018,Birrer2021} and are based on real images for the source and the lens. The full procedure is described in \cite{Rojas+2022} and can be summarized as follows: We pair Luminous Red Galaxies (LRGs) from the DES and source galaxies from the HST/HSC combined catalogue compiled by \cite{Canameras2020}, here the galaxies have the \textit{HST}/ACS F814W high-resolution \citep{Leauthaud2007,Scoville2007,Koekemoer2007} and the colour information from Hyper Suprime Cam (HSC) ultra-deep stack images \citep{Aihara2018}. We modelled the mass of the systems as a Singular Isothermal Ellipsoid (SIE), which has the following parameters: 
the Einstein radius ($\theta_E$), Position Angle, the axis ratio and the central position. The Einstein radius was calculated using the lens and source redshifts and the lens velocity dispersion. We inferred the lens galaxy redshift and velocity dispersion using a K-Nearest-Neighbors (KNN) algorithm, based on the assumption that galaxies with similar $gri$ magnitudes will also have similar redshifts and velocity dispersion, this procedure is explained in detail in \cite{Rojas+2022}. The rest of the parameters are derived by fitting an elliptical S\'ersic profile to the DES $r$-band image of the LRG. From this mass model we calculate the deflection of light, and trace rays back onto the source plane. The position of the source is randomly selected within a squeare that encloses the caustic curves. This process is done on an 0.03\arcsec pixel grid and then downsampled and PSF matched to the DES cutout of the LRG, and the flux scaled to the DES zero point. This simulated arc is then added to the DES LRG image to create a simulated lens.



The first data set is the "Bright simulations". This sample contains a selection of simulations used to train the Convolutional Neural Network (CNN) in \cite{Rojas+2022}, with the addition of smaller Einstein radius simulations in the range $0.8 \arcsec < \theta_E < 1.2 \arcsec$ that were not used to train the neural network in that work. In this dataset, the magnitude of the sources is boosted by one magnitude brighter than the observed HSC sources. This gives a population of bright lensing features designed for the CNN to easily learn the properties of strong lenses. These simulations should be the easiest for experts of identifying as strong lenses.

The second data is the "Default simulations". These were created with the same procedure as in \cite{Rojas+2022}, but without boosting the magnitude of the source. In this set of simulations we expect the systems to be harder to classify, since the signal-to-noise ratio (SNR) will be lower and the sources will stand out less brightly relative to the lensing galaxies.

Additionally to these two simulated sets of lenses we added the eleven lenses from the Sloan Lens ACS Survey
\citep[SLACS,][]{Auger2009} in the field of view of DES. These are spectroscopically confirmed, smaller Einstein radius systems, that are clear in Hubble Space Telescope imaging but represent a challenge to identify in ground-based resolution. Since small Einstein radius systems are expected to dominate in the real Universe \citep{Collett2015}, we include the SLACS sample to see if there is any chance to identify these systems with visual inspection of ground-based telescopes. Finally, four lens candidates from \citet{Rojas+2022} categorized with high scores in the "sure lens" list were also shown to the participant to see if our classifiers agreed with the authors of \citet{Rojas+2022}, we call this data set "R22 lenses". 

\subsubsection{Negative examples.}
We included three data sets with 150 objects each that contained non-lens examples. These samples test the classifiers' ability to reject non-lens systems. The first data set is a random sub-sample of the negative examples presented in the training set used to train the CNN in \cite{Rojas+2022}, we call this data set  "Non-lenses training set" and contains LRGs that were not used to create strong lens simulations~\footnote{it is not impossible that this sample contains a real lens, but it is statistically unlikely}. The second data set consisted of a random selection of objects that were classified by the CNN with scores near zero. We called this data set "CNN$_s$=0" and it contains stamps from the LRG selection that are highly improbable to contain any lens feature according to the CNN. The third data set is called "Non-lens simulations": we follow the same procedure as for strong lens simulations, but we disable the lensing deflections. Instead, we paint the source nearby to the LRG. This is designed to mimic a 'source in front of LRG' alignment, which is the most potential false positive for lens classification. This sample allows us to evaluate if classifiers are able to distinguish between real compact lenses and unlensed blue galaxies close to LRGs.

\subsubsection{Unlabeled data.}

We additionally included two data sets of unlabeled data, each set containing 300 objects. The first one contains the 300 best stamps graded by the CNN in \cite{Rojas+2022}, where 6 of them were flagged as "Maybe lens" in that work. The objective of this data set is to re-do the visual inspection and compare the classifications of the authors of \cite{Rojas+2022} with the participants in our experiment. We call this data set unlabeled as we do not have a confirmed classification of the objects displayed, and although this data set was previously inspected by another group we do not use this information as prior. The second set was created by selecting random objects with CNN scores distributed between 0.1 and 0.9 from the sample analysed in \cite{Rojas+2022}. The objective of this is both to see if CNN grades and expert grades are correlated and to see if a population of high-quality candidates are likely to be missed by CNNs.

\subsection{Participants.}
We asked all the participants to complete a google form requesting some basic and confidential information that we used to have a more deep analysis of this experiment. We asked three multiple choice questions. These questions and their options are: 

\begin{enumerate}
\item {\it How many years have you worked in the field of gravitational lensing?}  a. Less than one year, b. Between 1-4 years, c. Between 4-8 years, d. Between 8-12 years, e. More than 12 years. 
\item {\it What is your research status?} a. Master student, b. Ph.D. student, c.Postdoc, d.Professor/lecturer/similar, e. Amateur enthusiast\footnote{For better understanding we call this group "Citizen scientist" here.}. 
\item {\it How confident do you feel classifying lens systems?} a.Very confident, b. Confident, c. A bit confident, d. Not confident. 
\end{enumerate}

This information was asked so that we could search for correlations in the performance of the classifiers. 

A total of 80 people filled in the google form and a total of 69,592 classifications were made in the project. Some of the users contributed only a small number of classifications and 15\% of them did not perform any classification - we discarded the classifications of such users. We included all classifications made by users that analyzed more than 25\% of the sample (370 objects). This cutoff leaves 55 classifiers, where 51\% of them finished the whole project. Then we have a total of 66,835 classifications, with an average of 45 classifications per object. 

The breakdown of the participants into the categories of academic position, years of experience, and confidence are listed in Table~\ref{tab:participants}. A point of interest from the information compiled at this point is the correlation between confidence and experience. As is expected, classifiers with more experience (either a higher academic status or years working in the field) feel more confident performing the task as we can see in Appendix~\ref{appendix-conf}.

\begin{table}
	\centering
	\caption{Number of participants split in the three different categories requested at the beginning of this experiment.}
	\label{tab:participants}
	\begin{tabular}{|c|c|} 
		\hline
		Academic status & N. of participants  \\
        \hline
		Professors & 16  \\
		Postdocs & 13  \\
		Ph.D. Students & 15  \\
		Master Students & 3 \\
		Citizen Scientist & 8 \\
		\hline
		Years of experience in the field &   \\
		\hline
		More than 12 & 11  \\
		8-12 & 8  \\
		4-8 & 10  \\
		1-4 & 17  \\
		Less than a year & 9  \\
        \hline
        Confidence &  \\
		\hline
		Very confident & 16  \\
		Confident & 22  \\
		A bit confident & 14  \\
		Not confident & 3  \\
        \hline
	\end{tabular}
\end{table}


\section{Results A: The Discovery of Lenses with expert visual inspection}
\label{sec:resultsA}
In this section we look at the performance of the ensemble of classifiers in identifying strong lenses.

\subsection{Scoring objects}\label{subsec:eval_of_subj}
To compute a score for each object classified by our users, we translate the four different options into numbers as follows: "Certain lens" = 1,  "Probable lens" = 2/3, "Probably not lens" = 1/3 and "Very unlikely" = 0. The score for each object is the mean value of all its classifications. This gives every object a score between 0 and 1: objects scoring 1 are universally considered to be a strong lens and objects scoring 0 are universally considered non-lenses. 

\begin{figure*}
   \centering
   \includegraphics[width=\linewidth]{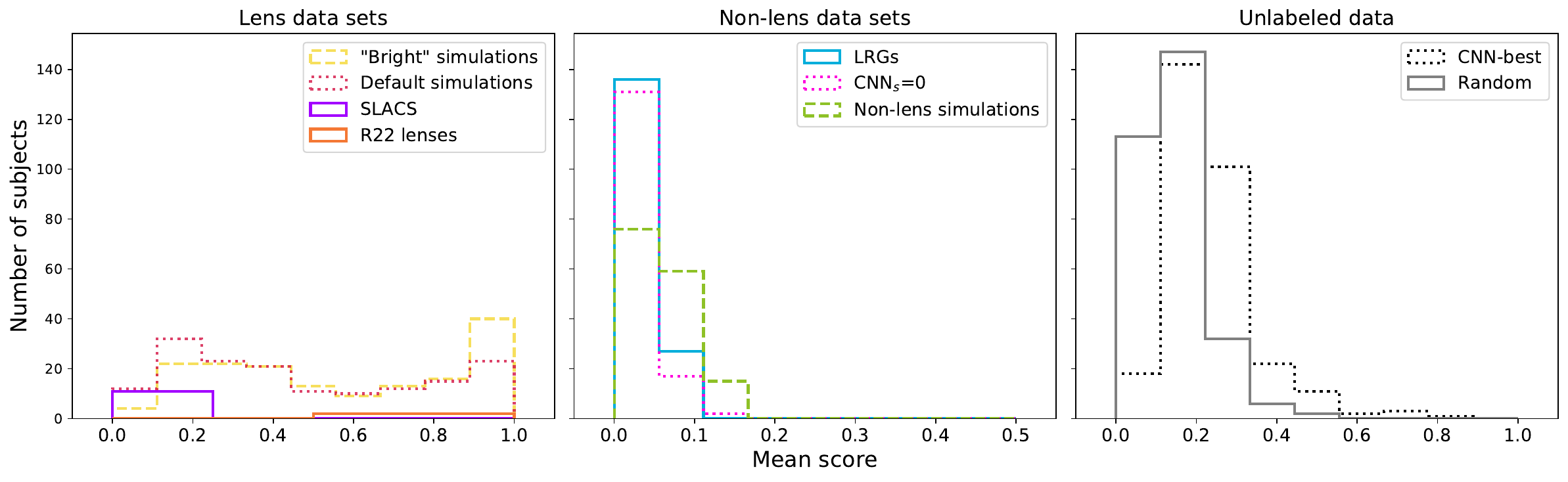}
    \caption{Mean score per object separated in each of the data sets presented in this work: lens examples (left panel), non-lens examples (middle panel), and unlabeled data (right panel). All the histograms have the same binning with the expectation of SLACS and R22 lenses data sets where the bin size is double for visualization purposes.}
    \label{fig:hist_meanscore}
\end{figure*}

In Fig.~\ref{fig:hist_meanscore} we present the mean score histograms split into the various data subsets. Overall we find that the scores of non-lenses are low and for many lenses the scores are high.

We investigated an alternative scoring system that up-weighted users with higher classification skill \citep{Marshall+2015}  but this had no impact on our results (See Appendix \ref{appendix-scoring}).


\subsection{Scores for images known to contain lenses.}
\begin{figure*}
   \centering
   \includegraphics[width=\linewidth]{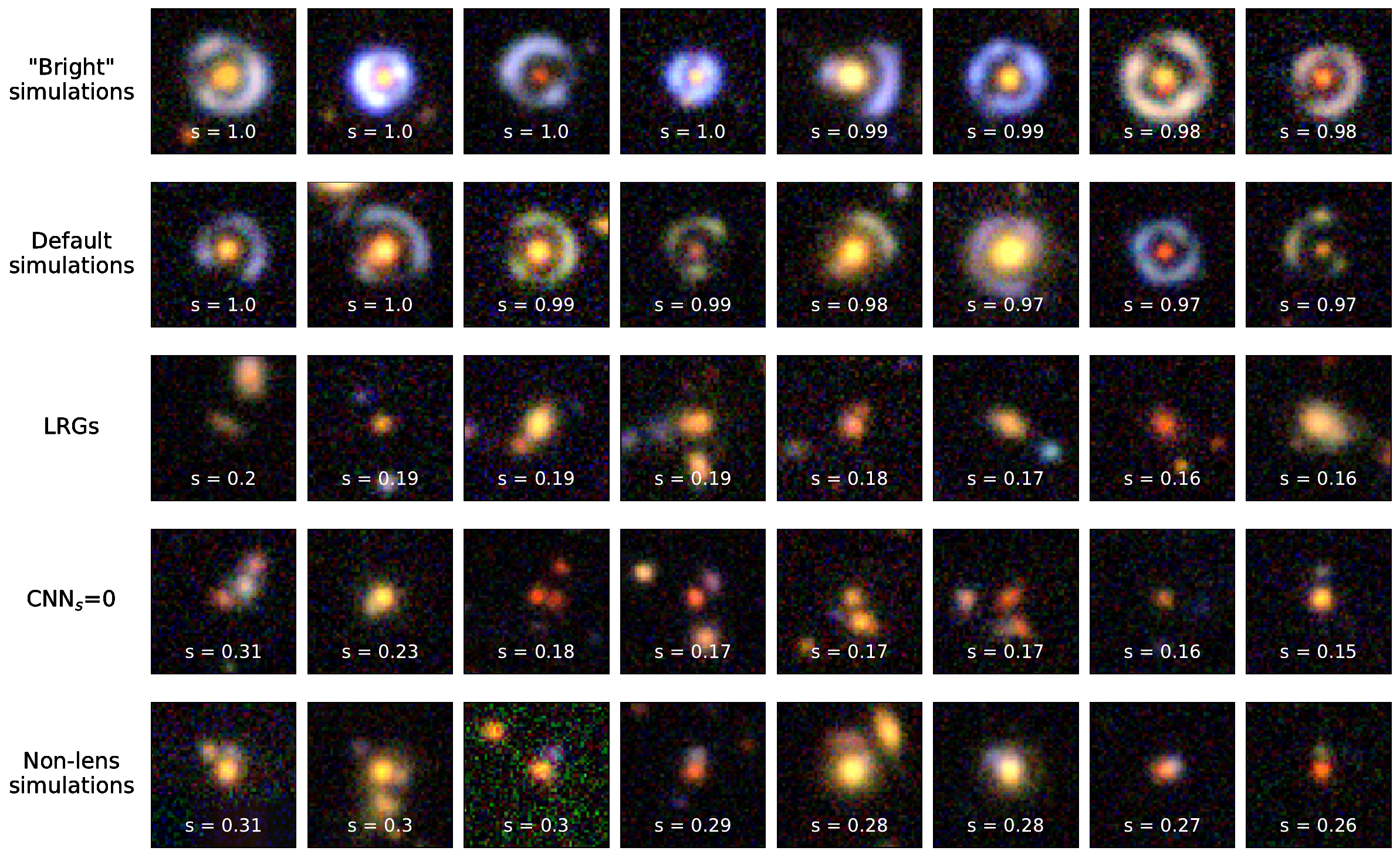}
    \caption{Examples of the eight objects with the highest mean score classified in the data sets: "Bright" simulations, Default simulations, LRGs, CNN$_{s}$=0 and non-lens simulations. The mean score is displayed at the bottom of each cutout.}
    \label{fig:mosaiclab}
\end{figure*}

The left panel in Fig.~\ref{fig:hist_meanscore} shows the distribution of scores for the four data sets of objects labelled as lenses.  Both simulated data sets have scores spanning the full range. This is not unsurprising: some of the simulations are bright arcs in textbook configurations whereas others are extremely faint or are not easily resolved from the lensing galaxy. It is not a surprise that the "Bright" simulation set contains a higher number of objects classified as lenses than in the "Default" one: the brightest arcs stand out more from the lenses. In Fig.~\ref{fig:mosaiclab} we present the 8 cutouts with the best scores for each of the lens simulation samples on the two top panels. 

\begin{figure*}
   \centering
   \includegraphics[width=\linewidth]{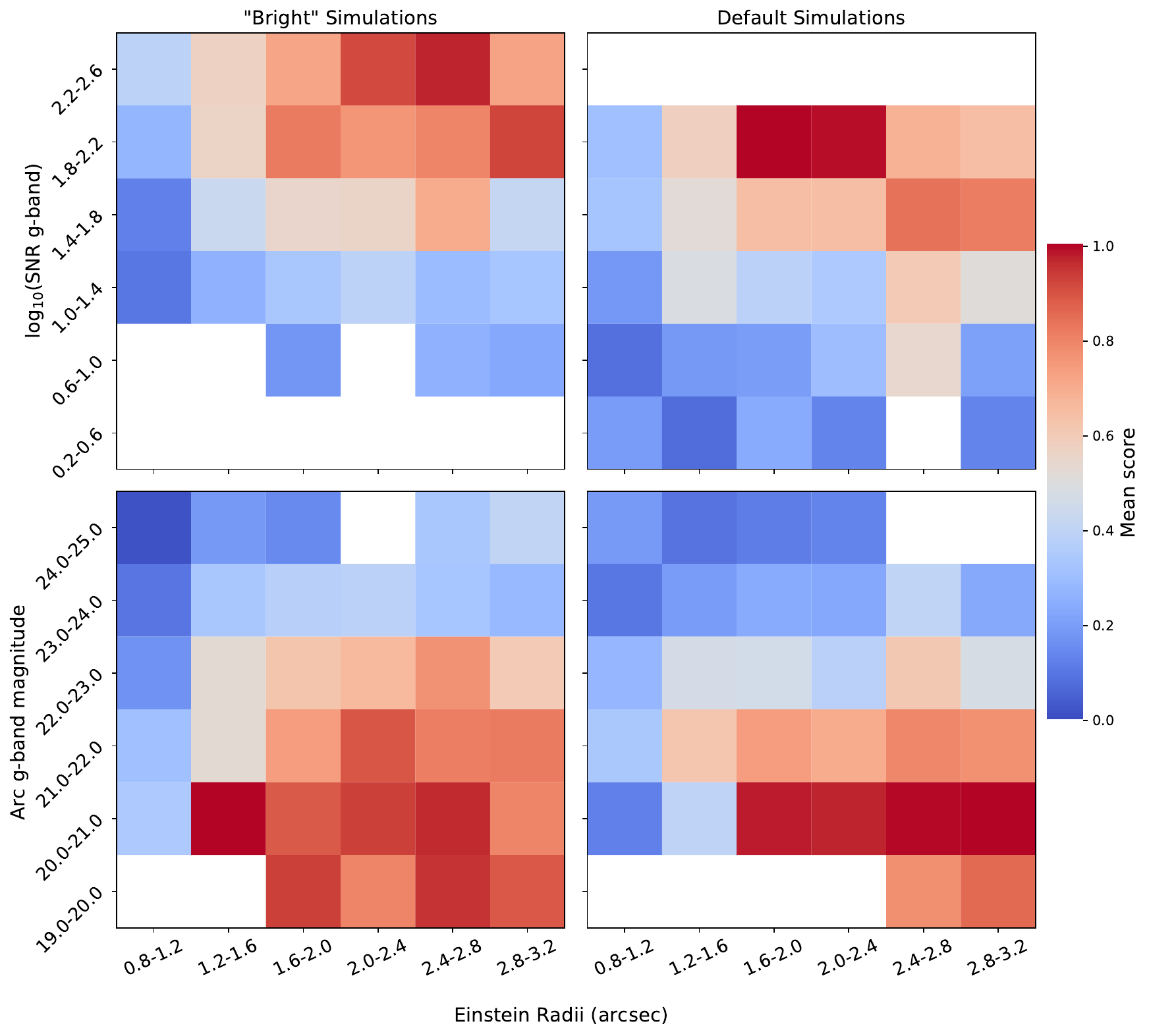}
    \caption{Heat maps with the average mean score per bins for the objects in the simulated lenses data sets. The left panels correspond to the "Bright" simulations, while the right panels to the "Default" simulations. In the top panels we display the logarithm of the SNR in g-band, while in the bottom panels we present the magnitude of the arc in the g-band. The color bar was selected with the purpose to more easily identify the bins where the simulations can be recognized as lens systems (in red) and where they are not identified (blue). }
    \label{fig:heatmap_bias}
\end{figure*}

The simulated lenses also give us an insight into the selection function of lens discovery with visual inspection. Since our sample is relatively small, we can only gain a coarse understanding of the selection function. In Fig.~\ref{fig:heatmap_bias} and Fig.~\ref{fig:heatmap_bias_std} we compare the recovery fraction of simulated lenses and its standard deviation as a function of signal-to-noise ratio in the g-band, the magnitude of the Arc in the g-band and the Einstein radius of the lens. Since all of our images are simulations of the approximately uniform depth Dark Energy Survey, the first two quantities are tightly correlated. It is clear from the heat maps of Fig.~\ref{fig:heatmap_bias} that there is a fairly sharp cutoff in recovery fraction for each quantity. Arcs fainter than $\sim$23rd magnitude (corresponding to a total SNR less than about 25), or with Einstein radius less than $\sim$1.2 \arcsec are not discoverable by human eye in DES-like imaging. On the other hand, from the standard deviation of the scores we do not see any trend that allows us to get further information.

\begin{figure*}
   \centering
   \includegraphics[width=\linewidth]{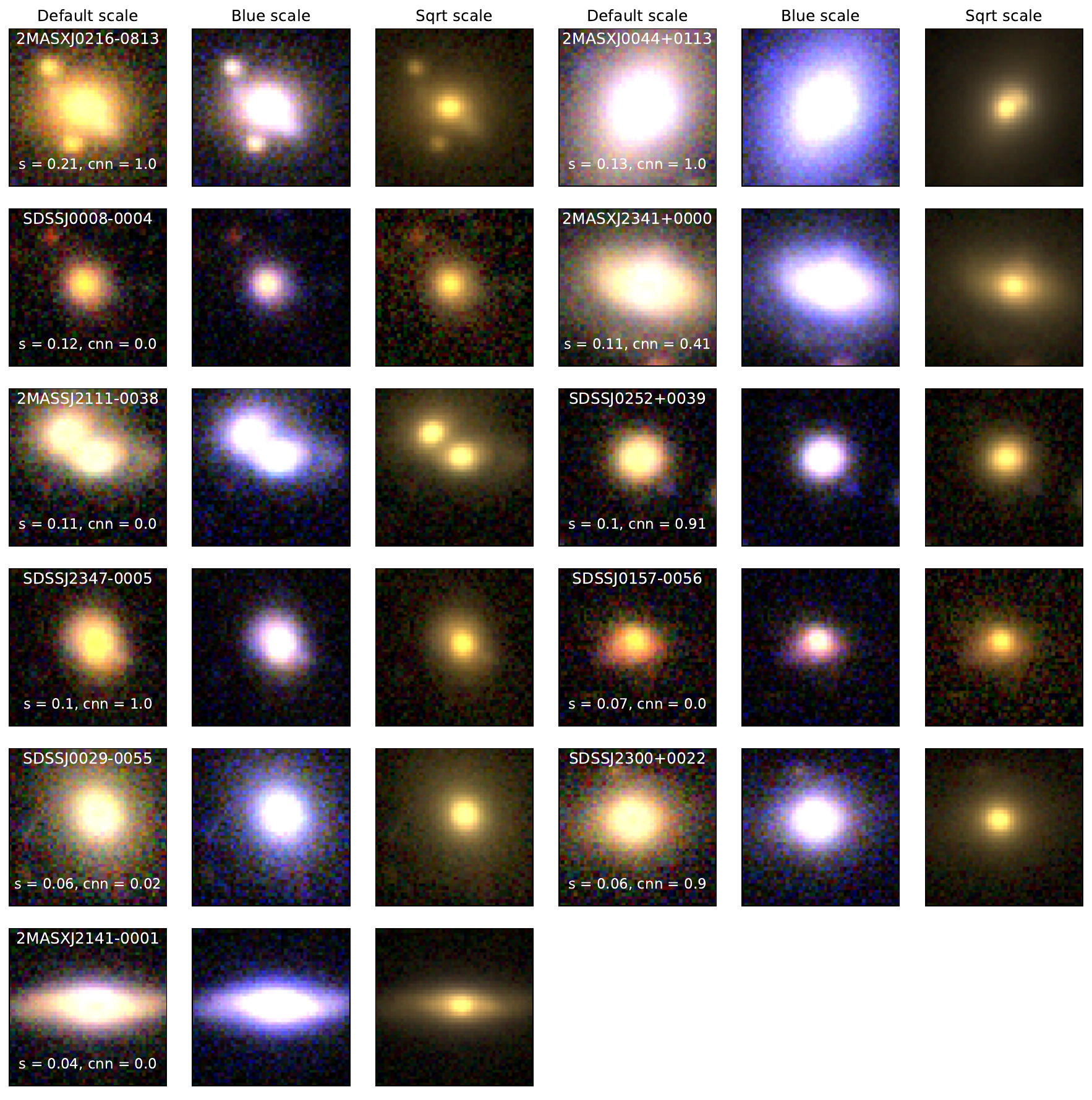}
    \caption{Mosaic of the SLACS sample in the DES. We displayed the cutouts of each system in the same three different scales presented in the experiment: Default, Blue and Sqrt. On the cutout with the Default scale we added on the top the name of the system and on the bottom the visual inspection mean score obtained in this search and the score given by the CNN trained in \protect\cite{Rojas+2022}.}
    \label{fig:mosaic_slacs}
\end{figure*}

In the same way, when we analyzed the SLACS sample we found that they were classified as "Non-lenses". The selection function of these systems drives them to have very bright lens galaxies and Einstein radii below 1.0 $\arcsec$ in most of the cases \citep{Dobler2008}. Given the results on similar simulated lenses, it is therefore not surprising that the human experts struggled to classify these systems as lenses. This is almost certainly because of the challenge to visually deblend the lens and source in DES imaging. The eleven SLACS systems in DES are shown in Fig.~\ref{fig:mosaic_slacs}, in the three different colour scales along with the score that they obtained from the visual inspection. 

\begin{figure}
   \centering
   \includegraphics[width=\linewidth]{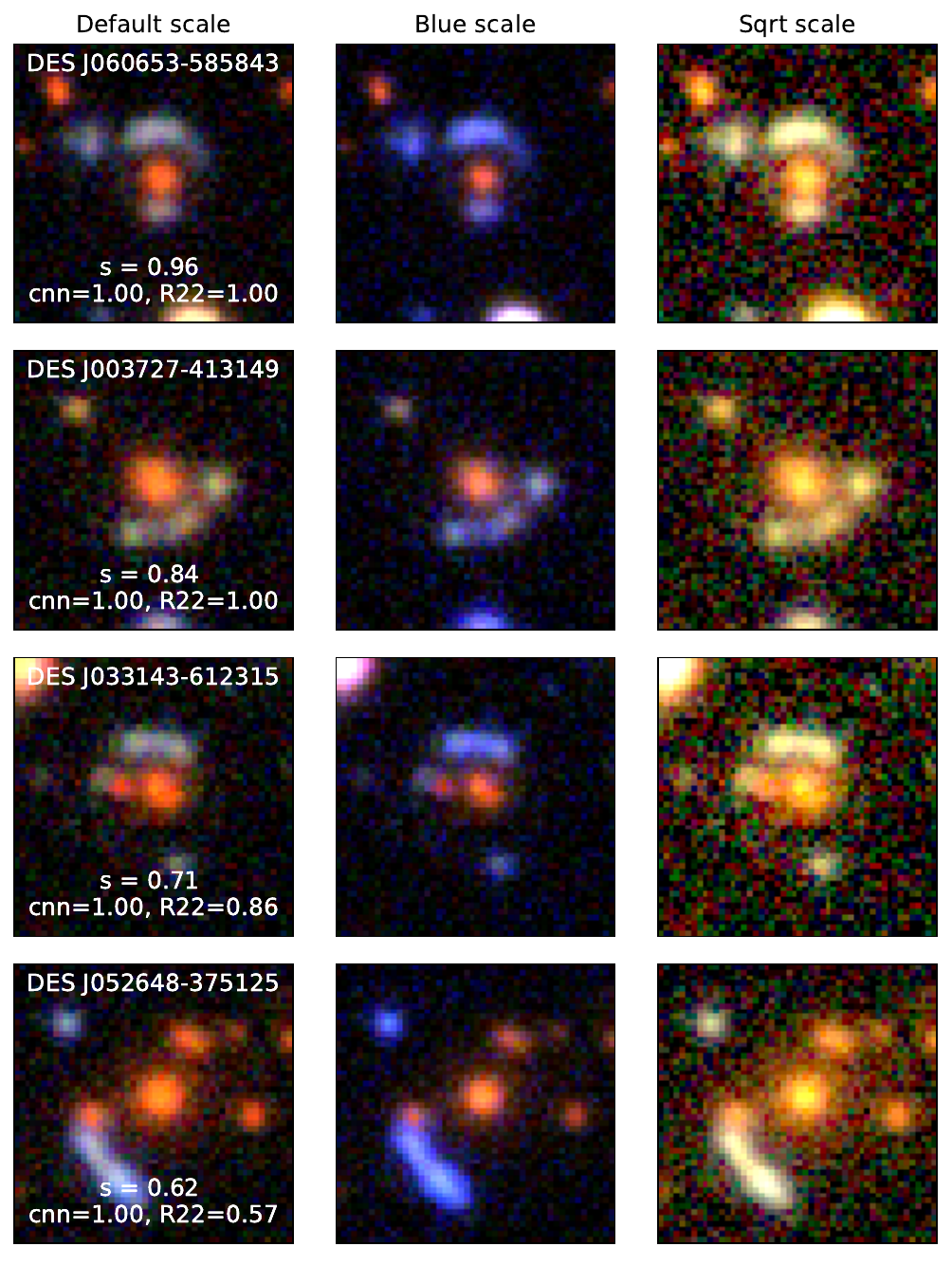}
    \caption{Mosaic of the four systems in \protect\cite{Rojas+2022} classified in their category "Sure lens" displayed in the three different scales presented in this experiment. On the cutout with the Default scale we added on the top the name of the system, and on the bottom the mean score (s) from our visual inspection, the CNN (cnn) and visual inspection score (R22) from \protect\cite{Rojas+2022}}
    \label{fig:mosaic_sl}
\end{figure}

On the other hand, the "R22 lenses" received scores between 0.6 and 1.0, i.e. the classifiers consider them to probably be lenses. These R22 lenses were discovered using a visual inspection of the same DES data, so it is not surprising that these lenses remain discoverable for our classifiers. Although, as we can see in Fig.~\ref{fig:mosaic_sl},  the visual inspection scores obtained in \cite{Rojas+2022} are somewhat different to those of our classifiers. We attribute this to human factors which we will discuss further in Sect.~\ref{subsec:score-numberofusers}. 


\subsection{Performance in negative examples.}

Our Non-lens examples are divided in three different data sets, the distribution of scores of their objects are shown in the middle panel of  Fig.~\ref{fig:hist_meanscore}. Here we see that most of them were classified with scores between 0 and 0.3, meaning they are very unlikely to be lenses. The "Non-lens simulations" (mimicking a chance non-lensing alignment) data set shows a broader distribution towards higher values these objects were potentially false positives, but they clearly do not particularly confuse our expert classifiers.

Even though most of the objects are correctly identified as non-lenses, in Fig.~\ref{fig:mosaiclab} we present the eight cutouts of each sample with higher scores. Here we can see that most of the objects in the "LRGs"  and "CNN$_{s}$=0" data sets have little blue or red-ish companions around the central galaxy that could be mistaken for signs of lensing. In the same way, the cutouts of the "Non-lens simulations" set can be easily mistaken by very compact (low Einstein radii) lens systems, producing that some users gave a higher score to these objects. Strictly speaking, the LRG and "CNN$_{s}$=0" sets could contain a lens, though the probability of this is $\ll1\%$. 

\subsection{Performance in unlabeled data.}\label{cnnvsvisualins}

We have two unlabeled data sets, "CNN-best" and "Random". Intuitively we should hope that the CNN-selected sample should contain more lens candidates than the random sample. All of the CNN-best lenses have been inspected in \cite{Rojas+2022}. In the right panel of Fig.~\ref{fig:hist_meanscore} we see that the CNN-best objects are distributed between 0.0 and 0.8 (0.5 in the case of the Random sample), but the peak of these distributions are around 0.2, with most of the objects classified as Non-lenses and only a few of them have a score above 0.5. 

\begin{figure*}
   \centering
   \includegraphics[width=\linewidth]{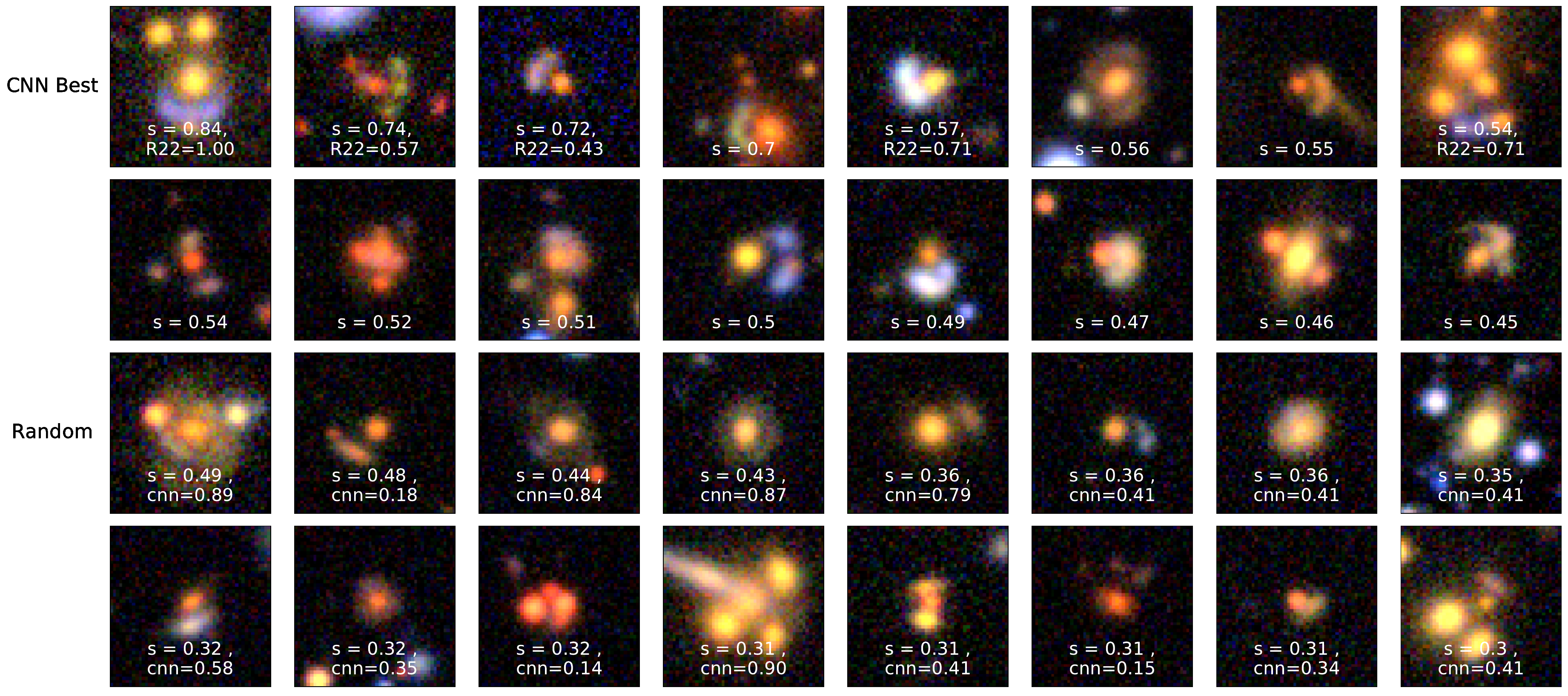}
    \caption{Examples of the 16 highly graded cutouts in the unlabeled data sets. The CNN-best data set is displayed in the two top panels, on the bottom of the cutouts we present the mean score (s) from this experiment, and visual inspection score (R22) from \citet{Rojas+2022} in case they were part of the final catalogs presented in that work. The CNN score for all these objects is CNN$ = 1.00$. The Random sample is shown in the two bottom panels, on the bottom of each cutout we display the  mean score (s) from this experiment, and the CNN score (cnn) from \citet{Rojas+2022}.}
    \label{fig:mosaic_unlab}
\end{figure*}

In Fig.~\ref{fig:mosaic_unlab} we present the 16 cutouts with the highest scores for each of the samples. In the "CNN-best" data set we find that five of the objects were also recognized as candidates in \cite{Rojas+2022}, all of them classified in their "Maybe lens" catalogue.  There are also seven objects with scores above 0.5 that were not classified as potential candidates in \cite{Rojas+2022}. In the case of the "Random" sample, none of the cutouts was classified as a potential lens candidate, although some get close to a score of 0.5. We can see some of them were highly graded by the CNN, this means that they went through the visual inspection steps in \citet{Rojas+2022} but were not selected as lenses by those authors.

\subsection{Comparison between CNN and visual inspection scores.}

\begin{figure}
   \centering
   \includegraphics[width=\linewidth]{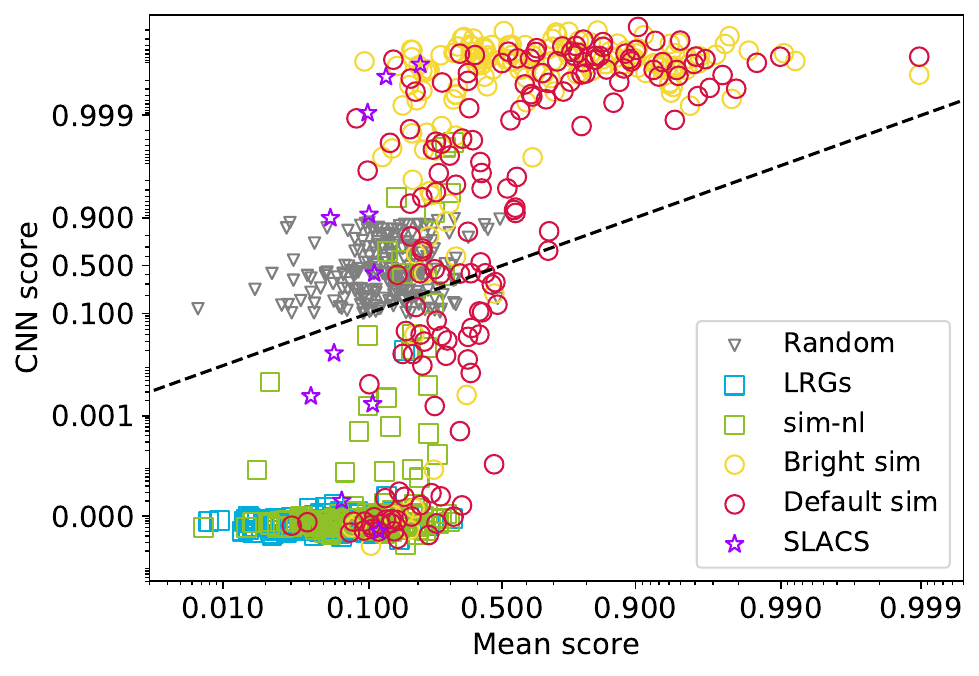}
    \caption{Comparison between the human expert scores from our classifiers and the CNN scores given by the model in \citet{Rojas+2022}. The dashed black line is the one-to-one line. Grey triangles are from the unlabeled 300 randomly drawn DES images. Squares are non-lens labelled data sets, in blue LRGs and in green the simulated non-lenses. Circles are simulated lenses, in yellow the bright data set and in red the default data set. Purple Stars are the SLACS lenses.}
    \label{fig:cnnscore}
\end{figure}

We compare the classification scores given by the CNN trained in \citet{Rojas+2022} with our visual inspection scores.  
Fig.~\ref{fig:cnnscore} shows the scatter graph of aggregated expert scores against the CNN scores of \citet{Rojas+2022}. These two scores are not strongly correlated, indicating  that the CNN and the experts are likely responding to different features in the images.

Since none of the objects in the Random data received a human score of 0.5 or more, we see no evidence of the CNN missing good candidates, however, this cannot be a definitive conclusion given the small sample size and the lack of correlation between the CNN and human scores. We cannot draw definitive conclusions from the CNN's strong performance on simulations, since the CNN was trained on simulations constructed in the same way.

\begin{figure*}
   \centering
   \includegraphics[width=\linewidth]{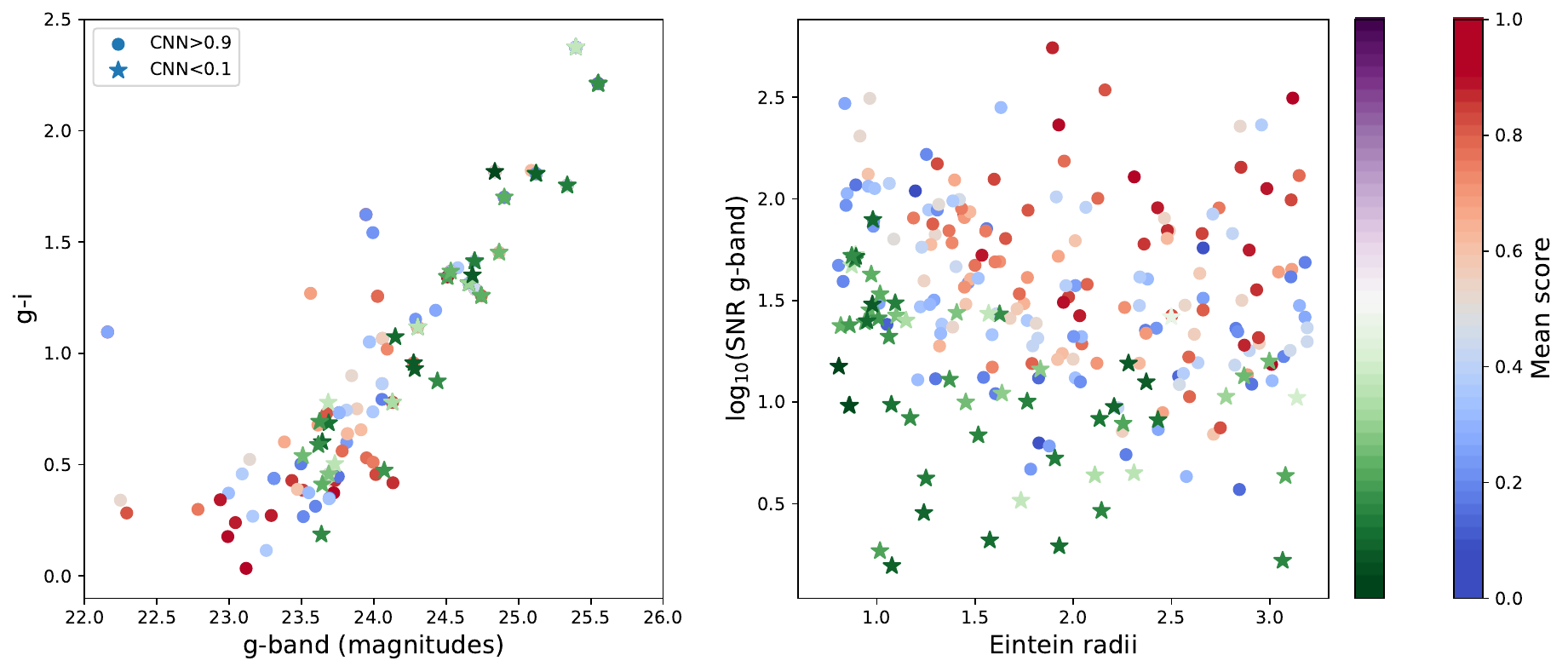}
    \caption{Scatter plots comparing source color (g-i) and g-band source magnitude (left) and Einstein radii and the logarithm of the SNR in g-band (right) for simulated lenses.  Circular blue-red markers have a CNN score above 0.9, Star shaped green-purple markers have a CNN score below 0.1. The markers are colored according to the mean visual inspection score.}
    \label{fig:tesnr}
\end{figure*}

In the non-lens data sets we see that most of the scores are well below 0.5, although a few non-lens simulations did manage to confuse the CNN. In a similar way, the CNN fails to recognize a subset of the lens simulations, often with a score even lower than given by the humans. In both simulated data sets the CNN correctly classifies around 1.6 times more images as lenses than humans, with the difference mostly coming from systems where the Einstein radii are below $1.2\arcsec$ 
(Fig.~\ref{fig:tesnr}). Similarly, some of the SLACS lenses obtained high scores by the CNN, despite being missed by the humans classifiers.

\cite{Jacobs2022} showed that, for CNN lens finders, parameters like color, PSF, occlusion and source magnitude play a major role in the CNN's scoring. We clearly see in that the source magnitude of plays a major role (see Fig~\ref{fig:tesnr}, left panel), with sources fainter than 24.5 magnitude not being detected by either the CNN and humans: fainter arcs are not detectable in DES-like imaging. We also see that the CNN gives higher scores to simulations with blue features, although it does reject other objects across the range of $g-i$ space. As such it appears that the CNN is correctly learning that most lensed sources are blue, rather than incorrectly assuming that most blue objects are lensed sources.

To understand cases where humans and the CNN gave contradictory scores, we compare the Einstein radii and SNR in g-band distributions for those systems (Fig.~\ref{fig:tesnr}, right panel). Several objects with high CNN scores but low visual inspection scores have small Einstein radii and/or low SNR, suggesting that the completeness of the CNN pushes further into this regime.
A mosaic with examples of the simulations with missmatched scores is presented in Fig.~\ref{fig:missmosaic}.



\section{Results B: Understanding expert classification}
\label{sec:resultsB}

In this section, we investigate how individual users performed when executing the classification task.

\subsection{How accurately do individuals classify our sample?}
\label{subsec:accuracy}
In order to evaluate the classification performed by each user we used the labeled data where we know the underlying truth: objects are either Lens or Non-lens. In our labeled category we find $58\%$ of all the classifications (38,432 in total). Knowing the true label of each object and the classification given by each user we can compute confusion matrices to compare user performance when classifying the objects.

\begin{figure}
   \centering
	\includegraphics[width=\linewidth]{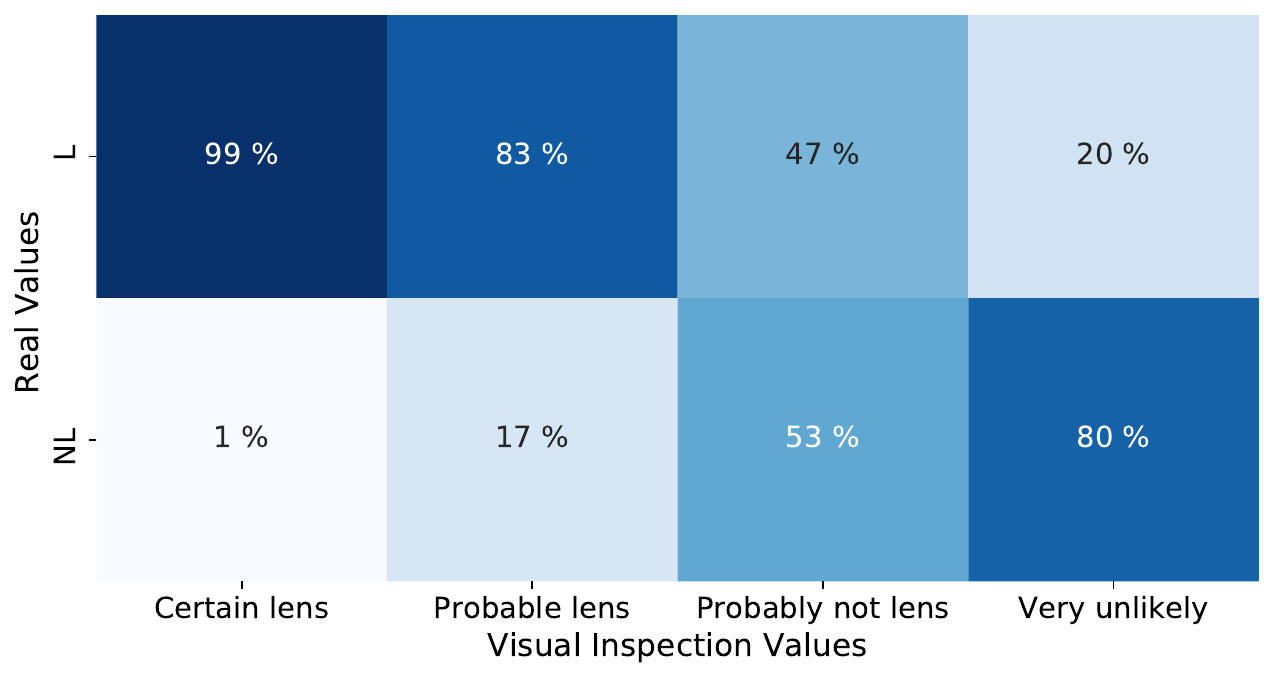}
    \caption{Confusion matrix of the four different classification options, "Certain lens", "Probable lens", "Probably not lens" and "Very unlikely" contrasted with the real labels L: lens and NL: No lens. The percentages shown are the number of lenses (non-lenses) classified in a determined option.}
    \label{fig:cfmatrix_4class}
\end{figure}

Aggregating all of the classifiers and classifications, we computed a confusion matrix displaying the four different classification options. In Fig.~\ref{fig:cfmatrix_4class}) we see the percentage of classifications that are in agreement or disagreement with the original label of the object. From here we can see that objects voted in the options "Certain lens", "Probable lens" and "Very unlikely" are in general well classified, achieving overall a $99\%$, $83\%$, and $80\%$ of objects correctly labelled. On the other hand the classification "Probably not lens" is evenly split between labelled lenses and labelled non-lenses. 

\begin{figure}
   \centering
	\includegraphics[width=\linewidth]{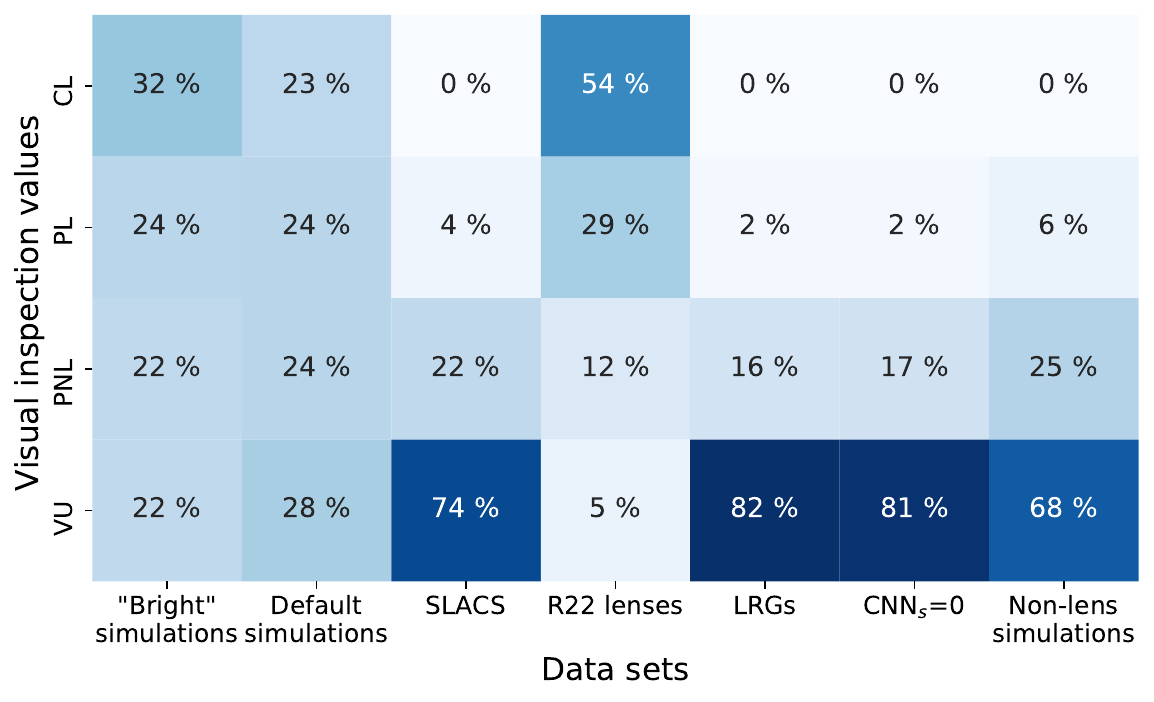}
    \caption{Confusion matrix of the seven different labeled data sets analyzed in this experiment contrasted with the classification options CL: Certain lens, PL: Probable lens, PNL: Probably not lens and VU: Very unlikely. The percentages represent the amount of classifications made in each category. 
 }
    \label{fig:cf_matrix_datasets}
\end{figure}

To see in more detail  the performance for each of the labelled data sets, we calculated a confusion matrix displaying the percentage of classifications according to the four different options for each data set. From Fig.~\ref{fig:cf_matrix_datasets} we can see that the three data sets labelled as "Non-lenses" obtained a very high percentage of votes in the option "Very Unlikely". However, there is confusion when classifying the simulated strong lens systems. SLACS lenses are mostly not recognized as lenses and only the four "R22 lenses" get a high amount of classifications in the categories "Certain lens" and "Probable lens". 

From the classifications of the labelled data, we compute confusion matrices for each user. In this section, we call objects Lenses if they are classified as either "Certain lens" or "Probable lens". Systems classified as  "Probably not lens" or "Very unlikely" are considered Non-lenses. In that way we build a $2\times2$ confusion matrix that will tell us what it is the probability that a user classifies an object as "Lens" or "Non-lens" given that the true label is "Lens" or "Non-lens", this means the true positive and true negative rates in the confusion matrix. 

\begin{figure}
   \centering
	\includegraphics[width=\linewidth]{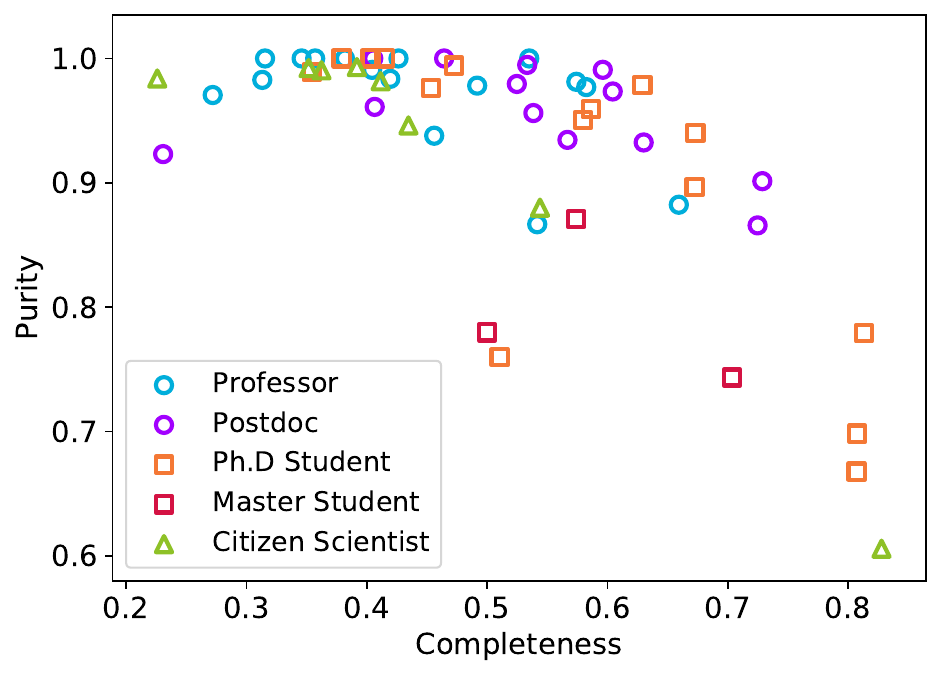}
    \caption{Completeness and purity percentages of each user when classifying our labelled data. 
}
    \label{fig:completnessvpurity}
\end{figure}

Following \cite{Marshall+2015} we plotted these probabilities in Fig.~\ref{fig:completnessvpurity}, where we can see that most of the classifiers have very low false positive rates. This makes these classifiers extremely good at identifying easy lenses, but there is a significant range in their ability to identify challenging lenses. Some classifiers manage completeness of  $\sim 60\%$ at high purity, whilst more pessimistic classifiers are identifying half as many lenses.

A handful of the classifiers are more optimistic, classifying more marginal systems as lenses. This comes at the cost of more false positives, which is not desirable in a real search where lenses are intrinsically rare.

\begin{figure}
   \centering
	\includegraphics[width=7cm]{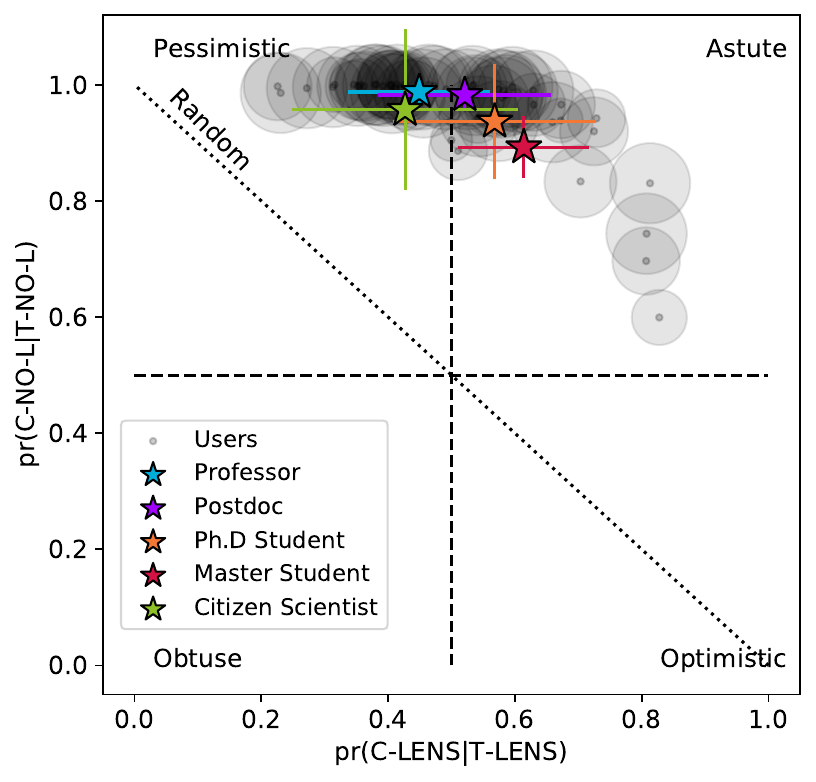}
	\includegraphics[width=7cm]{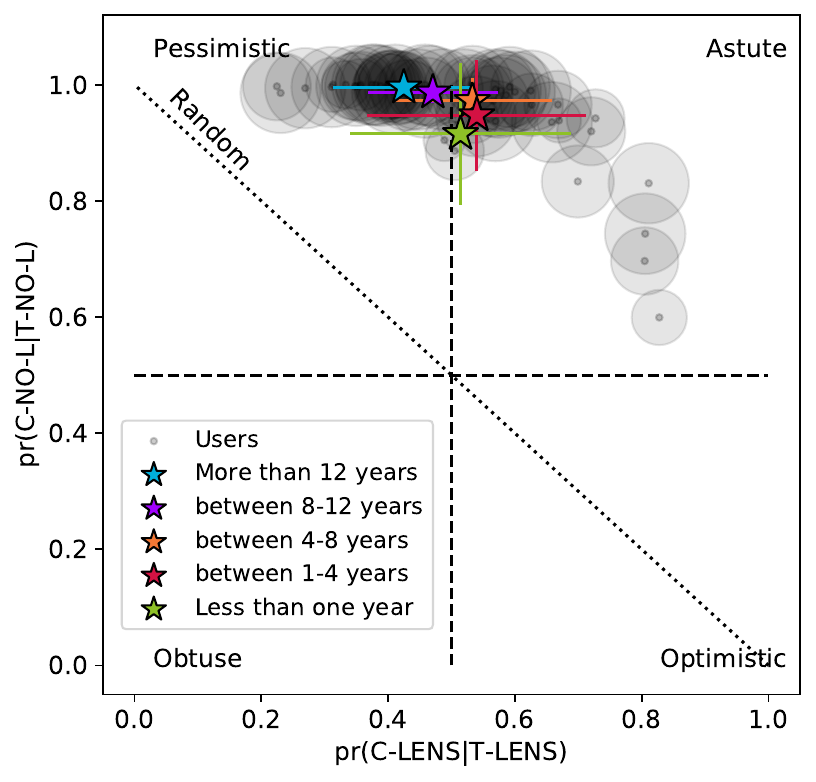}
	\includegraphics[width=7cm]{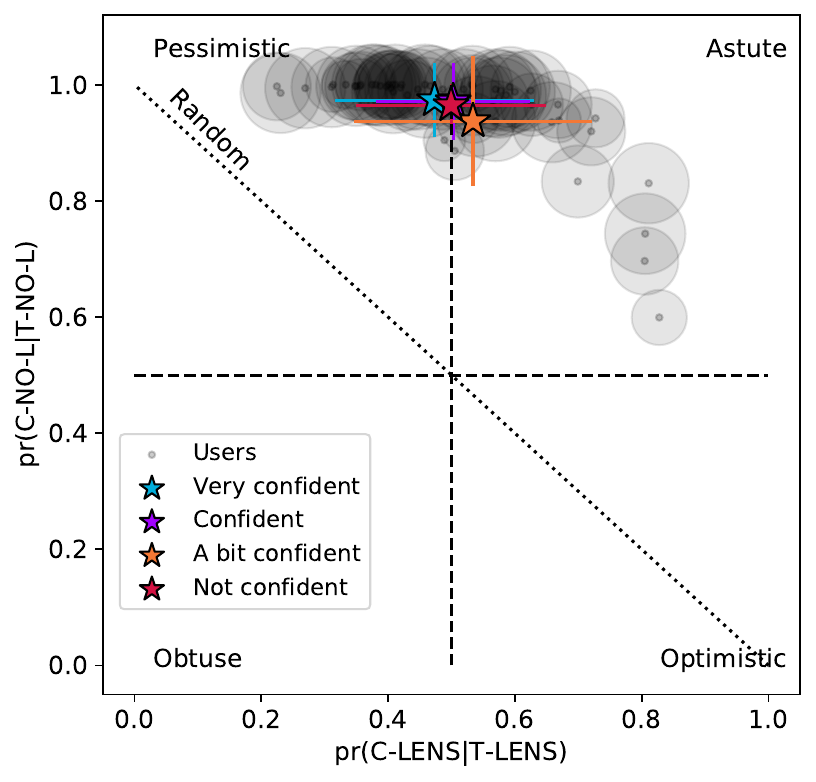}
    \caption{Probability that an object is classified as "Lens" given that it is a lens (true positive rate) vs. the probability that the object is classified as "Non-lens" given that it is not a lens (true negative rate). The values for each of the users are displayed with a black circle, whose size represents the amount of classification made by that user. The coloured stars represent the joint result of the different groups of classifiers presented in this work, separated by academic position (top panel), years of experience (middle panel) and confidence in performing the classification task (bottom panel).}
    \label{fig:classificationplot}
\end{figure}

Additionally, we computed the same $2\times2$ confusion matrix joining all the classifications of the users among the different groups separated by academic status, years of experience and confidence in performing the classification. This gives a confusion matrix for each grouping. To derive the error bars for these values, we use the standard deviation of the individual true positive and true negative rates of each user in the group. Fig.~\ref{fig:classificationplot} shows these results: overall the results are very similar. That is to say that, regardless of academic status, years of experience and confidence, each grouping produces a very similar average classification. There are very significant differences between individual users, but time in the field, academic position or reported confidence are not predictive of a user's classification skill.

\subsection{How reliable are individual classifications?}
\label{subsec:reliability}

To test the reliability of human classifications, we duplicated 105 objects in the sample. We randomly selected 15 objects from each of the following data sets: CNN-best, LRGs, $CNN_{s}$=$0$, Random, Default simulations, "Bright" simulations, and Non-lens simulations. The duplicate cutouts were shown at random points in the experiment~\footnote{Because of the way Zooniverse serves images, some users finished the classification task but continued to classify a small number of randomly drawn images. We also took these into account in assessing the reliability of classifications.}. 

Overall, we had a total of 3797 duplicated classifications, with $73\%$ graded the same as before. On the other hand, $15\%$ ($10\%$) received an upgrade (downgrade) of one point in the classification, this means that if for example an object was originally classified as "Probable lens" in the second time performing the classification the user classified the object as "Certain lens" in the upgrade case and as "Probably not lens" in the downgrade case. Only $1.3\%$ ($0.6\%$) of the classifications were upgraded (downgraded) by 2 points, and $0.13\%$ ($0.03\%$) by 3 points, which means changing completely the classification from "Certain lens" to "Very unlikely" or vice versa. 

\begin{figure}
   \centering
	\includegraphics[width=\linewidth]{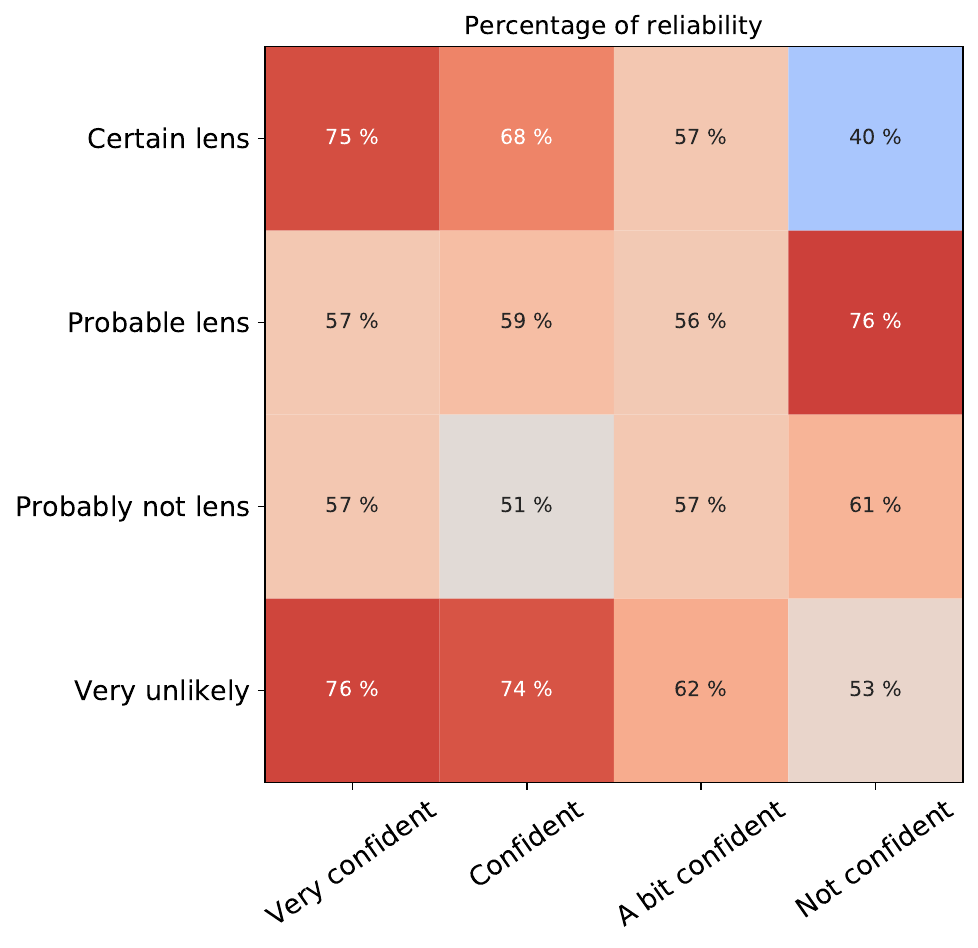}
    \caption{Reliability of repeat classifications by users, as a function of the classification and user reported confidence before starting the task. The percentages shown are the fraction of objects that have the same classification both times it was scored by a single user}
    \label{fig:reliability}
\end{figure}

These very low percentages for extreme cases are a very good sign that the users are not obtuse classifiers and are somehow confident about their classifications. When we break these results down by self-reported confidence (Fig.~\ref{fig:reliability}), we see that "Very confident" users perform relatively consistently, with a $\sim75\%$ of reliability in the two extreme classification options. On the other hand, the users that signed as "Not confident" hesitate more at the time to use the "Certain lens" option, reaching only $40\%$ of reliability in this class, but a $76\%$ of reliability in the option "Probable lens" shows that they are more comfortable with this more ambiguous selection. 

\subsection{How many classifiers are needed for a reliable score?}\label{subsec:score-numberofusers}

The classification of an object into any of the options is a personal and subjective opinion. Even with clear guidelines and examples, there will be disagreement among users, as the visual inspection work in \cite{Rojas+2022,Savary+22} showed. Typical strong lens searches have had a handful of expert classifiers \citep[e.g. 3 for][]{Jacobs2019A}. Given the individual expert variation seen in Sect.~\ref{subsec:accuracy} and the lack of reliability seen in Sect.~\ref{subsec:reliability}, it is to be expected that using a small number of experts can significantly bias final scores. To assess how significant this bias can be, we divide our classifications into random 'teams' of $n$ users. We computed new scores using only the classification of the team members. We did this for team sizes ranging from one to twenty participants randomly drawn from our users. 

\begin{figure}
   \centering
   \includegraphics[width=\linewidth]{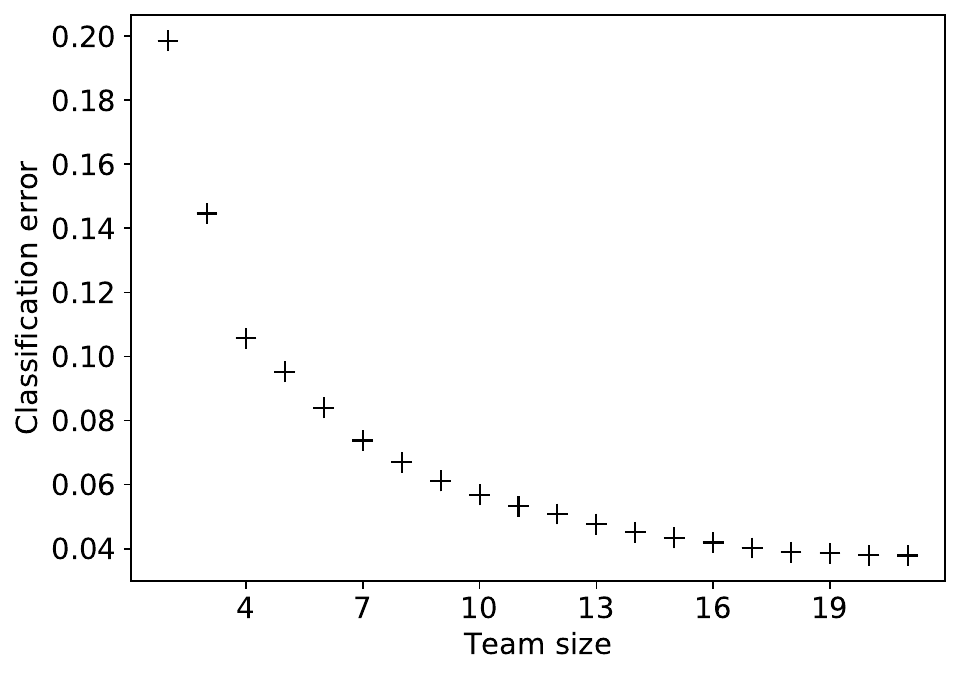}
    \caption{Comparison of the standard deviation of scores obtained from a subset (team) of users relative to the 'truth' from all classifiers. This figure shows that small teams are prone to inaccurate classifications  
    }
    \label{fig:numberusers}
\end{figure}

We compare the team scores with the final score from the entirety of our classifiers. We created 200 teams of n random users to recalculate the score for objects labelled as lenses. Fig.~\ref{fig:numberusers} shows how the team size affects the scoring of images. The standard deviation error is substantial for very small teams but decreases quickly up to a team size of $\sim 8$, which has a classification error of 0.07 per object. Above 8 users the classification error decreases much more slowly. Given that most objects are unambiguously classified as very unlikely (i.e. score of 0), these represent big differences in scores for the other objects: small teams are not very good at assessing the quality of a strong lens candidate.

The results in this section go some way to explaining why the scores of \citet{Rojas+2022} vary from those of our classifiers. There are not enough systems with overlapping data to draw firm conclusions (9 images), but the standard deviation is 0.08, as is expected for a team of 6 classifiers.


 Since teams will be most concerned about discovering marginal lenses, it is salient to focus on team accuracy when classifying lenses with marginal scores. Focusing on images with true scores between 0.3 and 0.8 the classification error grows substantially: it is 0.17 for a team of 2 classifiers, 0.12 for a team of 4 and 0.07 for a team of 8. A team of 6 classifiers is required to achieve an expected accuracy better than 0.1, 15 are needed for an accuracy better than 0.05.

\section{Conclusions}
\label{sec:conclusions}

This work has investigated how strong lensing experts visually inspect images of galaxy-scale strong lens candidates. We showed them a sample of 1489 mock and real images of the Dark Energy Survey and asked them to grade each image as either: Certain lens; Probable lens; Probably not a lens; or, Very unlikely to be a lens. With the resulting 66835 classifications we can now answer our initial questions.

\begin{itemize}
\item What are the properties of lensing systems that human experts identify reliably as lenses, and what do they miss?
\end{itemize}
Most gravitational lenses are reliably identified with an Einstein Radius greater than 1.2\arcsec and an arc g-band magnitude less than 23. This corresponds to roughly 1.2 times the seeing of the Dark Energy Survey and a g-band signal-to-noise of 25. Some lenses are discovered with fainter or smaller radius arcs. The Einstein radius cut-off is sharp with only a handful of very bright arcs discovered with Einstein radius of less than 1.2\arcsec. The flux cut-off is smoother: roughly twenty percent of lenses are recovered even with an arc signal-to-noise of between 4 and 10. 

\begin{itemize}
\item Do human experts confuse non-lenses for lenses?
\end{itemize}
For our labelled data, none of the non-lenses scored higher than 0.3. Our simulations do not include face-on spirals, or ring galaxies, but the experts had no problem rejecting chance alignments of blue sources close to LRGs. Our labelled sample had an almost equal split of lenses and non-lenses. Unless robotic candidate selection improves substantially, real lens searches will have far more non-lenses than lenses, even so, it seems that human experts are very good at discarding the kinds of non-lenses shown here. Follow-up campaigns should be confident that highly scored candidates are almost certainly lenses.

\begin{itemize}
\item How does expert classification depend on the experience and confidence of the experts?
\end{itemize}
We see substantial variation in the purity and completeness of individual classifiers, but there are no significant trends with experience, confidence, or academic position. None of these traits reliably predict the over-pessimism or over-optimism of some users.

\begin{itemize}
\item How reliable are individual classifications?
\end{itemize}
Classifications are not reliable when repeated. Even classifiers who self-report as 'very confident' do not grade candidates consistently. When reclassifying the same images, certain and very unlikely lenses are scored the same  roughly three-quarters of the time by confident and very confident classifiers, whereas probable and probably not lenses are only reproduced three-fifths of the time. Fewer than 2 percent of reclassified targets changed by more than one classification step. 

\begin{itemize}
\item When ranking lens candidates, what do the scores of teams of experts mean? How should lens searchers best build an expert team to classify their candidates?
\end{itemize}
Given the fact that classifications by a single expert are not reliable when repeated, it is not surprising that small teams make for poor classifiers. On a 0-1 scoring system, teams of 6 classifiers will produce results within 0.1 of the ensemble average of all users when classifying marginal systems. Senior classifiers are, on the whole, not better than junior classifiers so teams should classify independently and not defer to the opinions of senior faculty members. 

We found no correlation between CNN and human scores suggesting that CNNs are not trained to recognise the same features as human experts. Bigger samples are needed to assess if this is a problem for lens finding in future surveys.

A traditional search would use a small number of classifiers to grade a large number of images. To understand the human classification process, we have done the opposite. In the real Universe, real lenses are much rarer than our sample, so it is possible that our results do not perfectly scale to a search of a billion objects. However, if we assume that our discovery thresholds map onto searches of entire surveys our results suggest that previous forecasts are likely to be somewhat optimistic. The discovery signal-to-noise of our experts is broadly consistent with the assumptions of \citet{Collett2015}, however our experts recovered few lenses with Einstein radius less than 1.2\arcsec, which represents $\sim$40 percent of the forecasted DES population in \citet{Collett2015}. \citet{Collett2015} had assumed that users would be shown lens-subtracted images, such as in \citet{Sonnenfeld2018} and future work should investigate if expert inspection can recover even more lenses with such an approach.


\section*{Acknowledgements}
This work has received funding from the European Research Council (ERC) under the European Union's Horizon 2020 research and innovation programme (LensEra: grant agreement No 945536). TC is funded by the Royal Society through a University Research Fellowship.
For the purpose of open access, the authors have applied a Creative Commons Attribution (CC BY) licence to any Author Accepted Manuscript version arising.
DB is funded by a graduate studentship from UK Research and Innovation's STFC and the University of Portsmouth.
This work is also in part supported by the Swiss National Science Foundation (SNSF) and by the European Research Council (ERC) under the European Union's Horizon 2020 research and innovation program (COSMICLENS: grant agreement No 787886).
JHHC aknowledge the generosity of Eric and Wendy Schmidt by recommendation of the Schmidt Futures program.
FG acknowledges the support from grant PRIN MIUR 2017-20173ML3WW\_001.
RJ acknowledges the support from the research project grant "Understanding the Dynamic Universe" funded by the Knut and Alice Wallenberg Foundation under Dnr KAW 2018.0067.
S.S. thank the Max Planck Society for support through the Max Planck Research Group for SHS. This project has received funding from the European Research Council (ERC) under the European Unions Horizon 2020 research and innovation programme (LENSNOVA: grant agreement No 771776). This research is supported in part by the Excellence Cluster ORIGINS which is funded by the Deutsche Forschungsgemeinschaft (DFG, German Research Foundation) under Germany's Excellence Strategy  -- EXC-2094 -- 390783311. 
TD Thanks the support by an LSSTC Catalyst Fellowship awarded by LSST Corporation with funding from the John Templeton Foundation grant ID \#62192.
GM acknowledges funding from the European Union's Horizon 2020 research and innovation programme under the Marie Skłodowska-Curie grant agreement No MARACHAS - DLV-896778.
BD was supported in part by the NASA Astrophysics Theory Program under grant 80NSSC18K1014.
F.G.S. and G.V.C. Coordena\c{c}\~ao de Aperfei\c{c}oamento de Pessoal de Ensino Superior (CAPES) - Finance Code 001
LAU-L thanks CONACyT M\'exico for support under grants No. A1-S-17899, No. 286897, No. 297771, No. 304001; and the Instituto Avanzado de Cosmolog\'ia Collaboration.
JW was supported by the Science and Technology Facilities Council under grant number ST/P006760/1,  the DISCnet Centre for Doctoral Training in Data-Intensive Science.
JM acknowledges the support of the UK Science and Technology Facilities Council (STFC).

\section*{Data Availability}

The inclusion of a Data Availability Statement is a requirement for articles published in MNRAS. Data Availability Statements provide a standardised format for readers to understand the availability of data underlying the research results described in the article. The statement may refer to original data generated in the course of the study or to third-party data analysed in the article. The statement should describe and provide means of access, where possible, by linking to the data or providing the required accession numbers for the relevant databases or DOIs.



\bibliographystyle{mnras}
\bibliography{example} 

\begin{thebibliography}{}
\makeatletter
\relax
\def\mn@urlcharsother{\let\do\@makeother \do\$\do\&\do\#\do\^\do\_\do\%\do\~}
\def\mn@doi{\begingroup\mn@urlcharsother \@ifnextchar [ {\mn@doi@}
  {\mn@doi@[]}}
\def\mn@doi@[#1]#2{\def\@tempa{#1}\ifx\@tempa\@empty \href
  {http://dx.doi.org/#2} {doi:#2}\else \href {http://dx.doi.org/#2} {#1}\fi
  \endgroup}
\def\mn@eprint#1#2{\mn@eprint@#1:#2::\@nil}
\def\mn@eprint@arXiv#1{\href {http://arxiv.org/abs/#1} {{\tt arXiv:#1}}}
\def\mn@eprint@dblp#1{\href {http://dblp.uni-trier.de/rec/bibtex/#1.xml}
  {dblp:#1}}
\def\mn@eprint@#1:#2:#3:#4\@nil{\def\@tempa {#1}\def\@tempb {#2}\def\@tempc
  {#3}\ifx \@tempc \@empty \let \@tempc \@tempb \let \@tempb \@tempa \fi \ifx
  \@tempb \@empty \def\@tempb {arXiv}\fi \@ifundefined
  {mn@eprint@\@tempb}{\@tempb:\@tempc}{\expandafter \expandafter \csname
  mn@eprint@\@tempb\endcsname \expandafter{\@tempc}}}

\bibitem[\protect\citeauthoryear{{Aihara} et~al.,}{{Aihara}
  et~al.}{2018}]{Aihara2018}
{Aihara} H.,  et~al., 2018, \mn@doi [\pasj] {10.1093/pasj/psx066}, \href
  {https://ui.adsabs.harvard.edu/abs/2018PASJ...70S...4A} {70, S4}

\bibitem[\protect\citeauthoryear{{Auger}, {Treu}, {Bolton}, {Gavazzi},
  {Koopmans}, {Marshall}, {Bundy}  \& {Moustakas}}{{Auger}
  et~al.}{2009}]{Auger2009}
{Auger} M.~W.,  {Treu} T.,  {Bolton} A.~S.,  {Gavazzi} R.,  {Koopmans}
  L.~V.~E.,  {Marshall} P.~J.,  {Bundy} K.,   {Moustakas} L.~A.,  2009, \mn@doi
  [\apj] {10.1088/0004-637X/705/2/1099}, \href
  {https://ui.adsabs.harvard.edu/abs/2009ApJ...705.1099A} {705, 1099}

\bibitem[\protect\citeauthoryear{{Avestruz}, {Li}, {Zhu}, {Lightman}, {Collett}
   \& {Luo}}{{Avestruz} et~al.}{2019}]{Avestruz2019}
{Avestruz} C.,  {Li} N.,  {Zhu} H.,  {Lightman} M.,  {Collett} T.~E.,   {Luo}
  W.,  2019, \mn@doi [\apj] {10.3847/1538-4357/ab16d9}, \href
  {https://ui.adsabs.harvard.edu/abs/2019ApJ...877...58A} {877, 58}

\bibitem[\protect\citeauthoryear{{Birrer} \& {Amara}}{{Birrer} \&
  {Amara}}{2018}]{Birrer2018}
{Birrer} S.,  {Amara} A.,  2018, \mn@doi [Physics of the Dark Universe]
  {10.1016/j.dark.2018.11.002}, \href
  {https://ui.adsabs.harvard.edu/abs/2018PDU....22..189B} {22, 189}

\bibitem[\protect\citeauthoryear{Birrer et~al.,}{Birrer
  et~al.}{2021}]{Birrer2021}
Birrer S.,  et~al., 2021, \mn@doi [Journal of Open Source Software]
  {10.21105/joss.03283}, 6, 3283

\bibitem[\protect\citeauthoryear{{Bonvin} et~al.,}{{Bonvin}
  et~al.}{2017}]{Bonvin2017}
{Bonvin} V.,  et~al., 2017, \mn@doi [\mnras] {10.1093/mnras/stw3006}, \href
  {https://ui.adsabs.harvard.edu/abs/2017MNRAS.465.4914B} {465, 4914}

\bibitem[\protect\citeauthoryear{{Ca{\~n}ameras} et~al.,}{{Ca{\~n}ameras}
  et~al.}{2020}]{Canameras2020}
{Ca{\~n}ameras} R.,  et~al., 2020, \mn@doi [\aap]
  {10.1051/0004-6361/202038219}, \href
  {https://ui.adsabs.harvard.edu/abs/2020A&A...644A.163C} {644, A163}

\bibitem[\protect\citeauthoryear{{Christensen} et~al.,}{{Christensen}
  et~al.}{2012}]{Christensen2012}
{Christensen} L.,  et~al., 2012, \mn@doi [\mnras]
  {10.1111/j.1365-2966.2012.22007.x}, \href
  {https://ui.adsabs.harvard.edu/abs/2012MNRAS.427.1973C} {427, 1973}

\bibitem[\protect\citeauthoryear{{Collett}}{{Collett}}{2015}]{Collett2015}
{Collett} T.~E.,  2015, \mn@doi [\apj] {10.1088/0004-637X/811/1/20}, \href
  {https://ui.adsabs.harvard.edu/abs/2015ApJ...811...20C} {811, 20}

\bibitem[\protect\citeauthoryear{{Collett} \& {Auger}}{{Collett} \&
  {Auger}}{2014}]{Collett-Auger2014}
{Collett} T.~E.,  {Auger} M.~W.,  2014, \mn@doi [\mnras]
  {10.1093/mnras/stu1190}, \href
  {https://ui.adsabs.harvard.edu/abs/2014MNRAS.443..969C} {443, 969}

\bibitem[\protect\citeauthoryear{{Dobler}, {Keeton}, {Bolton}  \&
  {Burles}}{{Dobler} et~al.}{2008}]{Dobler2008}
{Dobler} G.,  {Keeton} C.~R.,  {Bolton} A.~S.,   {Burles} S.,  2008, \mn@doi
  [\apj] {10.1086/589958}, \href
  {https://ui.adsabs.harvard.edu/abs/2008ApJ...685...57D} {685, 57}

\bibitem[\protect\citeauthoryear{{Ebeling}, {Stockmann}, {Richard}, {Zabl},
  {Brammer}, {Toft}  \& {Man}}{{Ebeling} et~al.}{2018}]{Ebeling2018}
{Ebeling} H.,  {Stockmann} M.,  {Richard} J.,  {Zabl} J.,  {Brammer} G.,
  {Toft} S.,   {Man} A.,  2018, \mn@doi [\apjl] {10.3847/2041-8213/aa9fee},
  \href {https://ui.adsabs.harvard.edu/abs/2018ApJ...852L...7E} {852, L7}

\bibitem[\protect\citeauthoryear{{Flaugher} et~al.,}{{Flaugher}
  et~al.}{2015}]{Flaugher2015}
{Flaugher} B.,  et~al., 2015, \mn@doi [\aj] {10.1088/0004-6256/150/5/150},
  \href {https://ui.adsabs.harvard.edu/abs/2015AJ....150..150F} {150, 150}

\bibitem[\protect\citeauthoryear{{Gilman}, {Birrer}  \& {Treu}}{{Gilman}
  et~al.}{2020}]{Gilman2020}
{Gilman} D.,  {Birrer} S.,   {Treu} T.,  2020, \mn@doi [\aap]
  {10.1051/0004-6361/202038829}, \href
  {https://ui.adsabs.harvard.edu/abs/2020A&A...642A.194G} {642, A194}

\bibitem[\protect\citeauthoryear{{Honscheid} \& {DePoy}}{{Honscheid} \&
  {DePoy}}{2008}]{Honscheid2008}
{Honscheid} K.,  {DePoy} D.~L.,  2008, arXiv e-prints, \href
  {https://ui.adsabs.harvard.edu/abs/2008arXiv0810.3600H} {p. arXiv:0810.3600}

\bibitem[\protect\citeauthoryear{{Jacobs}, {Glazebrook}, {Collett}, {More}  \&
  {McCarthy}}{{Jacobs} et~al.}{2017}]{Jacobs2017}
{Jacobs} C.,  {Glazebrook} K.,  {Collett} T.,  {More} A.,   {McCarthy} C.,
  2017, \mn@doi [\mnras] {10.1093/mnras/stx1492}, \href
  {https://ui.adsabs.harvard.edu/abs/2017MNRAS.471..167J} {471, 167}

\bibitem[\protect\citeauthoryear{{Jacobs} et~al.,}{{Jacobs}
  et~al.}{2019a}]{Jacobs2019B}
{Jacobs} C.,  et~al., 2019a, \mn@doi [\apjs] {10.3847/1538-4365/ab26b6}, \href
  {https://ui.adsabs.harvard.edu/abs/2019ApJS..243...17J} {243, 17}

\bibitem[\protect\citeauthoryear{{Jacobs} et~al.,}{{Jacobs}
  et~al.}{2019b}]{Jacobs2019A}
{Jacobs} C.,  et~al., 2019b, \mn@doi [\mnras] {10.1093/mnras/stz272}, \href
  {https://ui.adsabs.harvard.edu/abs/2019MNRAS.484.5330J} {484, 5330}

\bibitem[\protect\citeauthoryear{{Jacobs}, {Glazebrook}, {Qin}  \&
  {Collett}}{{Jacobs} et~al.}{2022}]{Jacobs2022}
{Jacobs} C.,  {Glazebrook} K.,  {Qin} A.~K.,   {Collett} T.,  2022, \mn@doi
  [Astronomy and Computing] {10.1016/j.ascom.2021.100535}, \href
  {https://ui.adsabs.harvard.edu/abs/2022A&C....3800535J} {38, 100535}

\bibitem[\protect\citeauthoryear{{Jim{\'e}nez-Vicente}, {Mediavilla},
  {Kochanek}  \& {Mu{\~n}oz}}{{Jim{\'e}nez-Vicente}
  et~al.}{2015}]{Jimenez-Vicente2015a}
{Jim{\'e}nez-Vicente} J.,  {Mediavilla} E.,  {Kochanek} C.~S.,   {Mu{\~n}oz}
  J.~A.,  2015, \mn@doi [\apj] {10.1088/0004-637X/806/2/251}, \href
  {https://ui.adsabs.harvard.edu/abs/2015ApJ...806..251J} {806, 251}

\bibitem[\protect\citeauthoryear{{Koekemoer} et~al.,}{{Koekemoer}
  et~al.}{2007}]{Koekemoer2007}
{Koekemoer} A.~M.,  et~al., 2007, \mn@doi [\apjs] {10.1086/520086}, \href
  {https://ui.adsabs.harvard.edu/abs/2007ApJS..172..196K} {172, 196}

\bibitem[\protect\citeauthoryear{{Lanusse}, {Ma}, {Li}, {Collett}, {Li},
  {Ravanbakhsh}, {Mandelbaum}  \& {P{\'o}czos}}{{Lanusse}
  et~al.}{2018}]{Lanusse2018}
{Lanusse} F.,  {Ma} Q.,  {Li} N.,  {Collett} T.~E.,  {Li} C.-L.,  {Ravanbakhsh}
  S.,  {Mandelbaum} R.,   {P{\'o}czos} B.,  2018, \mn@doi [\mnras]
  {10.1093/mnras/stx1665}, \href
  {https://ui.adsabs.harvard.edu/abs/2018MNRAS.473.3895L} {473, 3895}

\bibitem[\protect\citeauthoryear{{Leauthaud} et~al.,}{{Leauthaud}
  et~al.}{2007}]{Leauthaud2007}
{Leauthaud} A.,  et~al., 2007, \mn@doi [\apjs] {10.1086/516598}, \href
  {https://ui.adsabs.harvard.edu/abs/2007ApJS..172..219L} {172, 219}

\bibitem[\protect\citeauthoryear{Marshall et~al.,}{Marshall
  et~al.}{2015}]{Marshall+2015}
Marshall P.~J.,  et~al., 2015, \mn@doi [Monthly Notices of the Royal
  Astronomical Society] {10.1093/mnras/stv2009}, 455, 1171

\bibitem[\protect\citeauthoryear{{Nierenberg} et~al.,}{{Nierenberg}
  et~al.}{2017}]{Nierenberg2017}
{Nierenberg} A.~M.,  et~al., 2017, \mn@doi [\mnras] {10.1093/mnras/stx1400},
  \href {https://ui.adsabs.harvard.edu/abs/2017MNRAS.471.2224N} {471, 2224}

\bibitem[\protect\citeauthoryear{{Oguri}, {Taruya}, {Suto}  \&
  {Turner}}{{Oguri} et~al.}{2002}]{Oguri2002}
{Oguri} M.,  {Taruya} A.,  {Suto} Y.,   {Turner} E.~L.,  2002, \mn@doi [\apj]
  {10.1086/339064}, \href
  {https://ui.adsabs.harvard.edu/abs/2002ApJ...568..488O} {568, 488}

\bibitem[\protect\citeauthoryear{{Petrillo} et~al.,}{{Petrillo}
  et~al.}{2017}]{Petrillo2017}
{Petrillo} C.~E.,  et~al., 2017, \mn@doi [\mnras] {10.1093/mnras/stx2052},
  \href {https://ui.adsabs.harvard.edu/abs/2017MNRAS.472.1129P} {472, 1129}

\bibitem[\protect\citeauthoryear{{Petrillo} et~al.,}{{Petrillo}
  et~al.}{2019}]{Petrillo2019}
{Petrillo} C.~E.,  et~al., 2019, \mn@doi [\mnras] {10.1093/mnras/stz189}, \href
  {https://ui.adsabs.harvard.edu/abs/2019MNRAS.484.3879P} {484, 3879}

\bibitem[\protect\citeauthoryear{{Rojas} et~al.,}{{Rojas}
  et~al.}{2022}]{Rojas+2022}
{Rojas} K.,  et~al., 2022, \mn@doi [\aap] {10.1051/0004-6361/202142119}, \href
  {https://ui.adsabs.harvard.edu/abs/2022A&A...668A..73R} {668, A73}

\bibitem[\protect\citeauthoryear{{Savary} et~al.,}{{Savary}
  et~al.}{2022}]{Savary+22}
{Savary} E.,  et~al., 2022, \mn@doi [\aap] {10.1051/0004-6361/202142505}, \href
  {https://ui.adsabs.harvard.edu/abs/2022A&A...666A...1S} {666, A1}

\bibitem[\protect\citeauthoryear{{Scoville} et~al.,}{{Scoville}
  et~al.}{2007}]{Scoville2007}
{Scoville} N.,  et~al., 2007, \mn@doi [\apjs] {10.1086/516580}, \href
  {https://ui.adsabs.harvard.edu/abs/2007ApJS..172...38S} {172, 38}

\bibitem[\protect\citeauthoryear{{Shu} et~al.,}{{Shu} et~al.}{2016}]{ShuY2016}
{Shu} Y.,  et~al., 2016, \mn@doi [\apj] {10.3847/1538-4357/833/2/264}, \href
  {https://ui.adsabs.harvard.edu/abs/2016ApJ...833..264S} {833, 264}

\bibitem[\protect\citeauthoryear{{Shu} et~al.,}{{Shu} et~al.}{2022}]{ShuX2022}
{Shu} X.,  et~al., 2022, \mn@doi [\apj] {10.3847/1538-4357/ac3de5}, \href
  {https://ui.adsabs.harvard.edu/abs/2022ApJ...926..155S} {926, 155}

\bibitem[\protect\citeauthoryear{{Sonnenfeld} et~al.,}{{Sonnenfeld}
  et~al.}{2018}]{Sonnenfeld2018}
{Sonnenfeld} A.,  et~al., 2018, \mn@doi [\pasj] {10.1093/pasj/psx062}, \href
  {https://ui.adsabs.harvard.edu/abs/2018PASJ...70S..29S} {70, S29}

\bibitem[\protect\citeauthoryear{{Stark} et~al.,}{{Stark}
  et~al.}{2015}]{Stark2015}
{Stark} D.~P.,  et~al., 2015, \mn@doi [\mnras] {10.1093/mnras/stv1907}, \href
  {https://ui.adsabs.harvard.edu/abs/2015MNRAS.454.1393S} {454, 1393}

\bibitem[\protect\citeauthoryear{{Tran} et~al.,}{{Tran}
  et~al.}{2022}]{Tran+2022}
{Tran} K.-V.~H.,  et~al., 2022, arXiv e-prints, \href
  {https://ui.adsabs.harvard.edu/abs/2022arXiv220505307T} {p. arXiv:2205.05307}

\bibitem[\protect\citeauthoryear{{Vegetti}, {Koopmans}, {Bolton}, {Treu}  \&
  {Gavazzi}}{{Vegetti} et~al.}{2010}]{Vegetti2010}
{Vegetti} S.,  {Koopmans} L.~V.~E.,  {Bolton} A.,  {Treu} T.,   {Gavazzi} R.,
  2010, \mn@doi [\mnras] {10.1111/j.1365-2966.2010.16865.x}, \href
  {https://ui.adsabs.harvard.edu/abs/2010MNRAS.408.1969V} {408, 1969}

\bibitem[\protect\citeauthoryear{{Wong} et~al.,}{{Wong}
  et~al.}{2020}]{Wong2020}
{Wong} K.~C.,  et~al., 2020, \mn@doi [\mnras] {10.1093/mnras/stz3094}, \href
  {https://ui.adsabs.harvard.edu/abs/2020MNRAS.498.1420W} {498, 1420}

\makeatother
\end{thebibliography}


\section*{Affiliations}
\noindent
{\it
$^{1}$ Institute of Cosmology and Gravitation, University of Portsmouth, Burnaby Rd, Portsmouth PO1 3FX, UK.\\
$^{2}$Kavli Institute for Particle Astrophysics and Cosmology and Department of Physics, Stanford University, Stanford, CA 94305, USA.\\
$^{3}$SLAC National Accelerator Laboratory, Menlo Park, CA, 94025.\\
$^{4}$Department of Physics and Astronomy, Stony Brook University, Stony Brook, NY 11794, USA.\\
$^{5}$Fermi National Accelerator Laboratory, P. O. Box 500, Batavia, IL 60510, USA. \\
$^{6}$Department of Astronomy and Astrophysics, University of Chicago, Chicago, IL 60637, USA.\\
$^{7}$Department of Physics and Astronomy, Lehman College of the CUNY, Bronx, NY 10468, USA. \\
$^{8}$Department of Astrophysics, American Museum of Natural History, Central Park West and 79th Street, NY 10024-5192, USA.\\
$^{9}$Institute of Physics, Laboratory of Astrophysics, Ecole Polytechnique F\'ed\'erale de Lausanne (EPFL), Observatoire de Sauverny, 1290 Versoix, Switzerland.\\
$^{10}$Instituto de Física de Cantabria. (CSIC-UC). Avda Los Castros s/n 39095 Santander. Spain.\\
$^{11}$University of Bologna, Department of Physics and Astronomy (DIFA), Via Gobetti 93/2, I-40129, Bologna, Italy. \\
$^{12}$INAF -- Osservatorio di Astrofisica e Scienza dello Spazio, via Gobetti 93/3 - 40129, Bologna - Italy.\\
$^{13}$Physics Department, University of Wisconsin-Madison, Madison, WI 53706, USA.\\
$^{14}$Oskar Klein Centre for Cosmoparticle Physics, Department of Physics, Stockholm University, Stockholm SE-106 91, Sweden.\\
$^{15}$Consejo Nacional de Ciencia y Tecnolog\'{i}a, Av.\ Insurgentes Sur 1582, Col.\ Cr\'{e}dito Constructor, Alc.\ Benito Ju\'{a}rez, CP 03940, M\'{e}xico. \\
$^{16}$Mesoamerican Centre for Theoretical Physics, Universidad Aut\'{o}noma de Chiapas,  Carretera Zapata Km. 4, Real del Bosque, 29040,\\
Tuxtla Guti\'{e}rrez, Chiapas, M\'{e}xico.\\
$^{17}$Max-Planck-Institut f\"ur Astrophysik, Karl-Schwarzschild Str. 1, 85748 Garching, Germany.\\
$^{18}$Technische Universit\"at M\"unchen, Physik Department, James-Franck Str. 1, 85748 Garching, Germany.\\
$^{19}$INAF -- Osservatorio Astronomico di Capodimonte, Salita Moiariello 16, 80131 - Napoli, Italy.\\
$^{20}$Instituto de Astronom\'ia, Observatorio Astron\'omico Nacional, Universidad Nacional Aut\'onoma de M\'exico, Apartado postal 106, C.P. 22800,\\
Ensenada, B.C., M\'exico.\\
$^{21}$Sub-department of Astrophysics, Denys Wilkinson Building, University of Oxford, Keble Road, Oxford, OX1 3RH, UK.\\
$^{22}$Department of Astrophysical Sciences, Princeton University, 4 Ivy Lane, Princeton, NJ 08544.\\
$^{23}$LSSTC Catalyst Fellow.\\
$^{24}$Department of Physics and Astronomy, School of Natural Sciences, University of Manchester, Oxford Rd, Manchester M13 9PL.\\
$^{25}$School of Physics \& Astronomy, University of Nottingham, University Park, Nottingham, NG7 2RD, UK.\\
$^{26}$Centre for Extragalactic Astronomy, Department of Physics, Durham University, Durham DH1 3LE, UK.\\
$^{27}$Institute for Computational Cosmology, Durham University, South Road, Durham DH1 3LE, UK.\\
$^{28}$Observat\'orio Nacional, Rua Gal. Jos\'e Cristino 77, Rio de Janeiro, RJ - 20921-400, Brazil.\\
$^{29}$OmegaLambdaTec GmbH, Lichtenbergstra{\ss}e 8, 85748 Garching.\\
$^{30}$Universit\"ats-Sternwarte, Fakult\"at für Physik, Ludwig-Maximilians Universit\"at M\"unchen, Scheinerstr. 1, 81679 M\"unchen, Germany.\\
$^{31}$Citizen Scientist, Zooniverse, Astrophysics Sub-department, University of Oxford, Keble Road, Oxford, OX1 3NP, UK.\\
$^{32}$Department of Physics and Astronomy, University of New Mexico, 210 Yale Blvd NE, Albuquerque, NM 87106, USA.\\
$^{33}$Centre for Astrophysics and Supercomputing, Swinburne University of Technology, PO Box 218, Hawthorn, VIC 3122, Australia. \\
$^{34}$University of Birmingham, School of Physics and Astronomy, Edgbaston, Birmingham, B15 2TT, UK. \\
$^{35}$Departamento de F\'isica, DCI, Campus Le\'on, Universidad de Guanajuato, 37150, Le\'on, Guanajuato, M\'exico.\\
$^{36}$Department of Physics, Benedictine University, 5700 College Road, Lisle, Illinois, 60532.\\
$^{37}$School of Physical Sciences, The Open University, Walton Hall, Milton Keynes, MK7 6AA, UK.\\
$^{38}$IRAP, Université de Toulouse, UPS, CNRS.\\
$^{39}$Institut f\"{u}r Theoretische Astrophysik, Zentrum f\"{u}r Astronomie, Heidelberg Universit\"{a}t, Albert-Ueberle-Str. 2, 69120, Heidelberg, Germany.\\
$^{40}$Instituto Sciety Lab, S\~ao Jos\'e dos Campos, SP, Brazil \\
$^{41}$Department of Physics, University of California, 1156 High St, Santa Cruz, CA 95064, USA.\\
$^{42}$School of Computing \& Communications, The Open University, Walton Hall, Milton Keynes MK7 6AA, UK.\\
$^{43}$Santa Cruz Institute of Particle Physics, University of California, 1156 High St, Santa Cruz, CA, 95064 USA.\\
$^{44}$The Inter-University Centre for Astronomy and Astrophysics (IUCAA), Post Bag 4, Ganeshkhind, Pune 411007, India.\\
$^{45}$Kavli Institute for the Physics and Mathematics of the Universe(IPMU), 5-1-5 Kashiwanoha, Kashiwa-shi, Chiba 277-8583, Japan.\\
}


\appendix

\section{Invitation}
\label{appendix-invitation}
We send an invitation to members of the strong lensing community, including LSST, DES, and Euclid strong lensing working groups and we extended the invitation to Space Warp citizen scientists. They received the following invitation:

\begin{quote}
"We would like to invite you to participate in a lens classification experiment with the goal of understanding how people in the field of gravitational lensing are performing when they do a visual classification task. The task will take about an hour of your time. Participants will be invited to coauthor the resulting paper.

The motivation of this experiment comes from the explosion of new lens systems discovered by the use of CNN and subsequent validation through visual inspection performed for each team. There seem to be lots of differences in the expert validation and we want to see if this can be understood and calibrated.

We expect that with this social experiment we can get key conclusions about our performance, and hope that in the future lens finders can benefit from this information. 
If you want to be part of this experiment please fill this quick google form first. The information requested here will help us to analyze the data, although your personal data (name, email and galaxy zoo username) will remain private.

And now you are welcome to classify ~1000 objects! If you follow this link\footnote{\url{ https://www.zooniverse.org/projects/krojas26/experts-visual-inspection-experiment/classify}}

You will see DES gri-color composite images of each object, each stamp has a size of 50x50 pixels (13"x13"). The same object is displayed in 3 different color-scales to help the recognition of features. The task is simple: you have to click on the option that better represents the object(s) in the image and go to the next.

You don't have the obligation to complete the classification all at once, this might take you a couple of hours. Your progress is recorded and you can come back anytime you want ;). For a successful analysis we hope you can commit to completing the classification of the whole data set, or at least a big portion of it.  

Please share this among your group, postdocs, phd students, masters students, etc, that work in the field of strong lensing. All are encouraged to participate as we want to test a broad variety of expertise, but please do not share it among people outside of the field." 
\end{quote}

\section{Image scalings}
\label{appendix-images}
 The "default" and "blue" composite images are scaled with an arcsinh stretch using \citep[{\tt HumVI},][]{Marshall+2015}. We tuned the $rgb$-scale parameters, the brightness (Q) and contrast ($\alpha$) for default (blue) images as follows: r-scale = 1.0 (0.51), g-scale = 1.2 (0.68), b-scale= 2.8 (3.12), Q = 1.0 (0.64), $\alpha$ = 0.03 (0.03). The third image is scaled using a square root image scaling. To ensure that the object of interest is not overshadowed by another brighter one we set minimum and maximum values for the pixels in the images. The minimum value is obtained using a square of $5 \times 5$ pixels in the corners of the $gri$-images and we select the minimum value among them. On the other hand, the maximum value is obtained by placing a box of $10 \times 10$ pixels in the middle of the three images and obtaining the mean among them. We use these two values to scale the three $gri$-images.

\section{Experience and Confidence}
\label{appendix-conf}
\begin{figure}
	\includegraphics[width=\columnwidth]{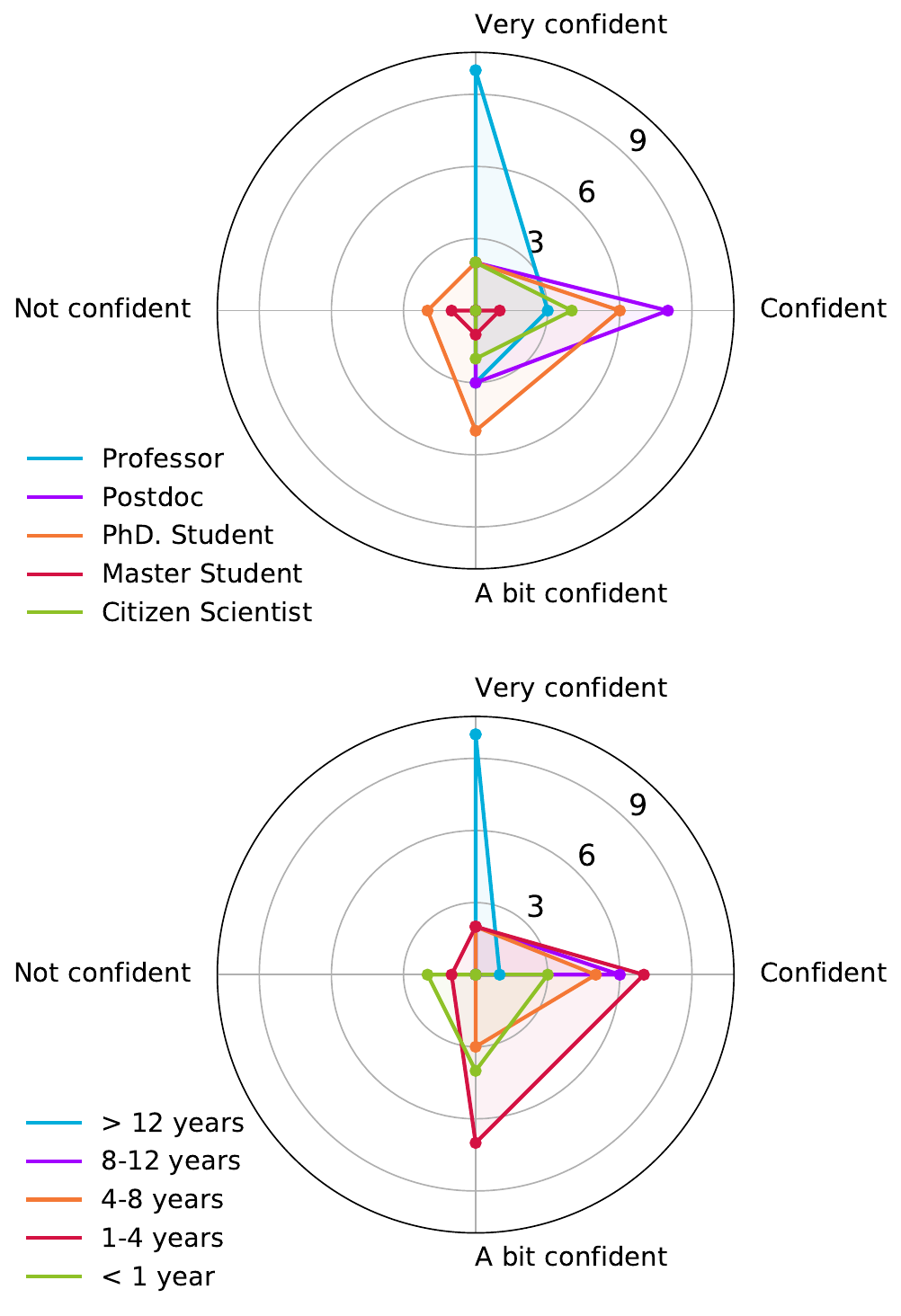}
    \caption{The radar charts show the level of confidence that users have to perform the classification task separately in academic status (top panel) and years of experience (bottom panel).}
    \label{fig:spiderplot}
\end{figure}

In Fig.~\ref{fig:spiderplot} we present the distribution of the levels of confidence compared to the academic status and years of experience in the field, according to the information provided by the classifiers when they subscribed to this experiment. From both plots, we can clearly see that at higher academic status or years of experience in the field, the classifiers feel more confident that they will perform a successful classification, while undergrad students and people with between 0-4 years of experience feel "not" or only a "bit confident".  

\section{Scoring system}
\label{appendix-scoring}
In addition, we wanted to explore the impact of weighting the classification of the users according to their percentage of correct classifications in each of the 4 options given. To obtain this percentage we computed a $2\times4$ confusion matrix similar to Fig.~\ref{fig:cfmatrix_4class} but this time for each user. Then we take the 4 relevant percentages: Classified as "Certain lens" or "Probable lens" given that it is a lens and "Probably not lens" or "Very unlikely" given that it is not a lens. Using these percentages as weight we then calculated a weighted mean score, this means that we multiply each score by the corresponding weight, sum them and divide by the sum of the weights. After weighting all the classifications we re-scale the mean score weighted between 0 and 1 as we did with the previous score. 

\begin{figure}
   \centering
	\includegraphics[width=7cm]{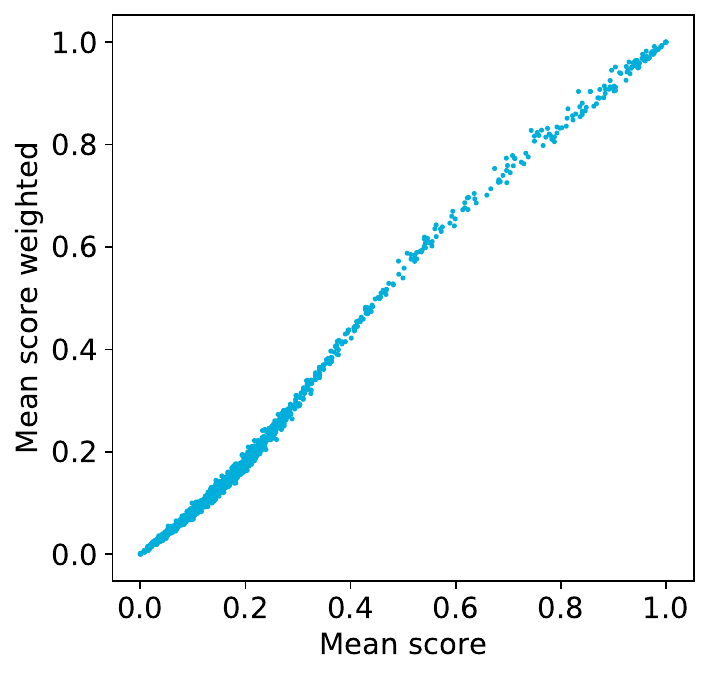}
    \caption{Scatter plot of the mean score compared with the mean score weighted by the corresponding percentage of successful classification in the options displayed.}
    \label{fig:scatterscore}
\end{figure}

We expected that weighting the mean score could provide a better score system because it will take into account the performance of the users, but most of the users performed in a pretty similar way. We can see in Fig.~\ref{fig:scatterscore} that the mean score and the mean score weighted are highly correlated. For this reason, we are going to continue our analysis using only the mean score and we are not going to explore the mean score weighted implications in further analysis. 

\section{Standard deviation}
\label{appendix-std}

\begin{figure*}
   \centering
   \includegraphics[width=\linewidth]{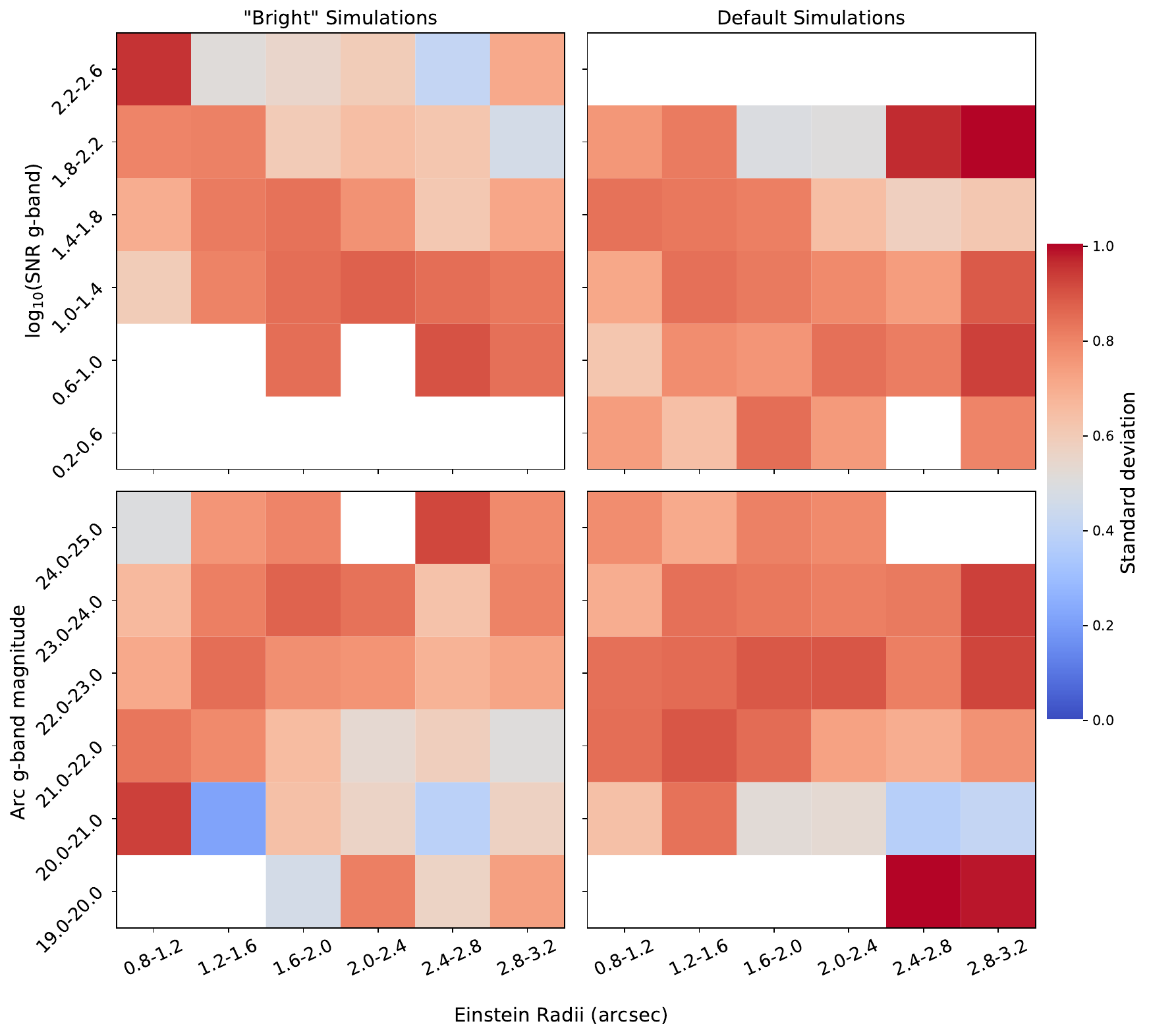}
    \caption{Heat maps with the average standard deviation of the mean score per bins for the objects in the simulated lenses data sets. The left panels correspond to the "Bright" simulations, while the right panels to the "Default" simulations. In the top panels we display the logarithm of the SNR in g-band, while in the bottom panels we present the magnitude of the arc in the g-band. }
    \label{fig:heatmap_bias_std}
\end{figure*}

We calculated the standard deviation of the scores given to each object in the experiment. With this information, we create a heat map  (Fig.~\ref{fig:heatmap_bias_std}) to compare the standard error of the recovery fraction of the simulated lenses data sets in response to the signal-to-noise ratio in the g-band, the magnitude of the arc in the g-band and the Einstein radius of the lens. In contrast with the heat maps with the average scores (Fig.~\ref{fig:heatmap_bias}), we do not see any trend in the standard deviation.

\section{Visual inspection vs CNN score}
\label{appendix-vicnn}
\begin{figure*}
   \centering
   \includegraphics[width=17cm]{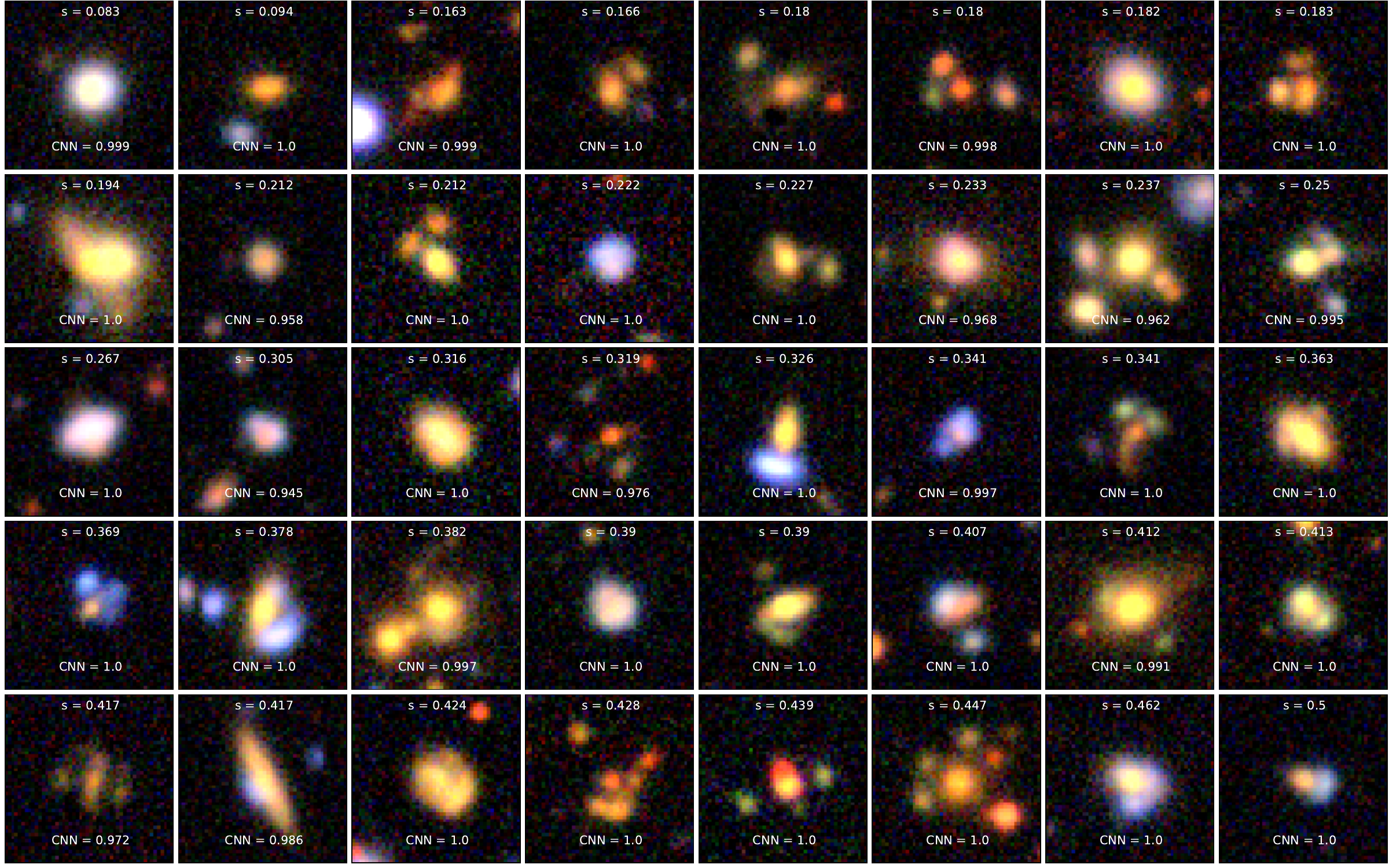}
   \\
   \vspace{0.5 cm}
   \includegraphics[width=17cm]{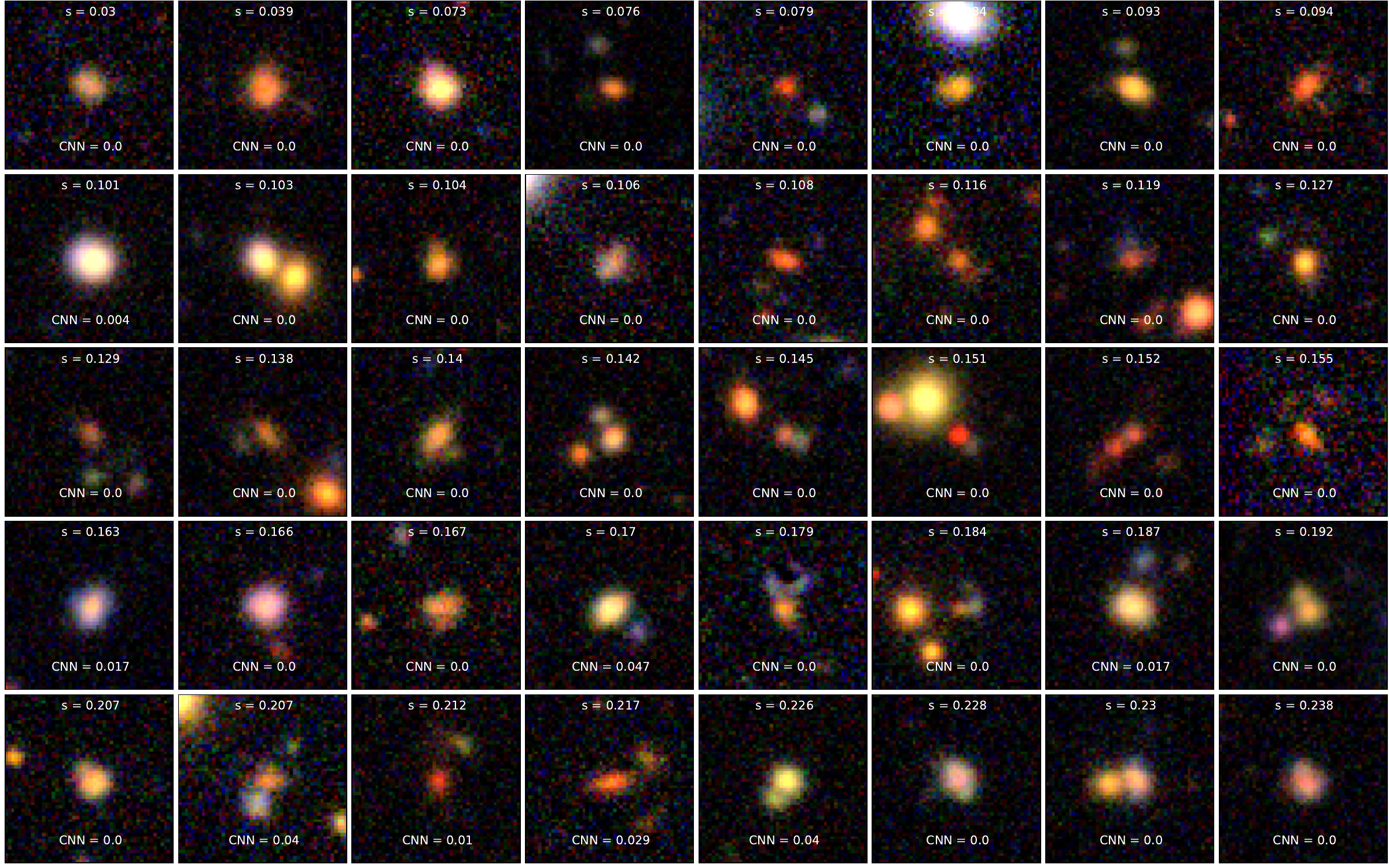}
    \caption{Mosaic with 80 examples of simulations that human visual inspectors graded with a score below 0.5. The first 40 are images where the humans and CNN disagree (CNN with a score above 0.9) and the other 40 show broad agreement between human and CNN (CNN score<0.1
    0)}
    \label{fig:missmosaic}
\end{figure*}

In Sect.~\ref{cnnvsvisualins} We compared the results of this experiment with the score given by the CNN trained by \cite{Rojas+2022}. In order to understand why the CNN is able to classify slightly more systems than human we displayed in Fig.~\ref{fig:missmosaic} examples of those simulations that the CNN graded as good candidates (CNN > 0.9) but human visual inspectors graded with a low score (s < 0.5), meaning they were classified as not good lens candidates. Additionally we explored the opposite CNN range of classification score (CNN < 0.1), where CNN and humans agree given that all simulations obtain s < 0.5, as can be seen in Fig.~\ref{fig:tesnr}.

\bsp	
\label{lastpage}
\end{document}